\documentclass[reprint,amsmath,amssymb,aps,superscriptaddress]{revtex4-2}

\usepackage{graphicx}
\usepackage{dcolumn}
\usepackage{bm}

\usepackage{booktabs}
\usepackage[table]{xcolor}

\usepackage[version=3]{mhchem}
\usepackage{setspace}
\usepackage{xfrac,nicefrac}

\usepackage{tabularray}
\UseTblrLibrary{booktabs}

\usepackage[justification=raggedright,singlelinecheck=false]{caption}
\usepackage[colorlinks=true,allcolors=blue]{hyperref}

\begin{document}

\title{Detection of MEMS Acoustics via Scanning Tunneling Microscopy}

\author{R.~J.~G.~Elbertse\textsuperscript{\S}}
\affiliation{Kavli Institute of Nanoscience, Department of Quantum Nanoscience, Delft University of Technology, Delft, The Netherlands}

\author{M.~Xu\textsuperscript{\S}}
\affiliation{Kavli Institute of Nanoscience, Department of Quantum Nanoscience, Delft University of Technology, Delft, The Netherlands}
\affiliation{Department of Precision and Microsystems Engineering, Delft University of Technology, Delft, The Netherlands}

\author{A.~Keşkekler}
\affiliation{Department of Precision and Microsystems Engineering, Delft University of Technology, Delft, The Netherlands}

\author{S.~Otte}
\affiliation{Kavli Institute of Nanoscience, Department of Quantum Nanoscience, Delft University of Technology, Delft, The Netherlands}
\email{a.f.otte@tudelft.nl} 

\author{R.~A.~Norte}
\affiliation{Department of Precision and Microsystems Engineering, Delft University of Technology, Delft, The Netherlands}
\affiliation{Kavli Institute of Nanoscience, Department of Quantum Nanoscience, Delft University of Technology, Delft, The Netherlands}
\email{r.a.norte@tudelft.nl} 

\begin{abstract}
Scanning tunneling microscopy (STM) and micro-electromechanical systems (MEMS) have traditionally addressed vastly different length scales—one resolving atoms, the other engineering macroscopic motion. Here we unite these two fields to perform minimally invasive-measurements of high-aspect-ratio MEMS resonators using the STM tip as both actuator and detector. Operating at cryogenic temperatures, we resolve acoustic modes of millimeter-scale, high-$Q$ membranes with picometer spatial precision, without making use of lasers or capacitive coupling. The tunneling junction introduces negligible back-action or heating, enabling direct access to the intrinsic dynamics of microgram-mass oscillators. In this work we explore three different measurement modalities, each offering unique advantages. Combined, they provide a pathway to quantum-level readout and exquisite high-precision measurements of forces, displacements, and pressures at cryogenic conditions. This technique provides a general platform for minimally-perturbative detection across a wide range of nanomechanical and quantum devices.
\end{abstract}
\maketitle

\vspace{-10pt}
\begin{center}
\footnotesize \textsuperscript{\S}These authors contributed equally to this work.\\
\footnotesize Corresponding authors: \texttt{a.f.otte@tudelft.nl}, \texttt{r.a.norte@tudelft.nl}
\end{center}
\vspace{6pt}

\thanks{Corresponding author: }

\section{Introduction}\label{sec1}

In recent decades, nano-membranes have facilitated breakthroughs in quantum science and precision sensing. Their thin geometries create minimal physical connection to external environments, which, combined with high tensile stresses, allows for dissipation dilution that achieves state-of-the-art mechanical isolation from surrounding environments. This isolation enables the probing of macroscopic quantum phenomena present in the membranes. Nano-membranes are widely used in quantum-limited sensing, where their relatively large surface areas make them ideal for sensing macroscopic forces such as pressure, acceleration, and Casimir effects \cite{Xu2025arXiv}. \\

Common interrogation methods for membrane acoustics includes optics~\cite{Thompson2008Nature,norte2016,Reinhardt2016PRX,tsaturyan2017ultracoherent} and electronics~\cite{andrews2014bidirectional,malcovati2018evolution,zawawi2020review}. However, these methods require strong interaction with the membrane's,  potentially introducing noise or perturbations via heating and charging, obscuring properties such as superconductivity or ferromagnetism. Recently, techniques using scanning force microscopy on graphene membranes has been presented \cite{Halg2020PRA}. While less invasive, such techniques still utilize lasers for the excitation and measurement of membrane resonances. Scanning microwave microscopy has been used to image acoustic modes of small silicon nitride membranes with microwave tips which are 50-550nm away from the membrane~\cite{xu2024imaging}. In contrast, here we use a scanning tunneling microscope (STM) tip to both excite and measure the resonances of (super)conducting membranes. The STM tip is uniquely capable of measuring the local density of states on surfaces with sub-nanometer precision via small tunneling currents. We show the capability to excite the membrane while simultaneously probing its oscillation amplitude using STM, achieving subatomic spatial precision with minimal invasiveness.

It is important to note that since 2012 there have been a number of studies~\cite{uder2018low, breitwieser2017, alyobi2020voltage, zan2012scanning, klimov2012electromechanical, xu2012atomic, zhao2013fabrication, breitwieser2014flipping, schoelz2015graphene, uder2017convenient} using scanning tunneling microscopy to approach free-standing graphene membranes of dimensions ranging from a few microns~\cite{uder2018low,alyobi2020voltage} to tens of microns~\cite{breitwieser2017}. These studies focused on manipulating wrinkling on these highly compliant membranes and manipulating their shapes, rather than engaging with their resonant acoustic modes. In contrast, our work enables the driving and measurement of acoustic modes in membranes that are approximately 100,000 times more massive than any other object manipulated with scanning tunneling microscopy to date. In doing so, we connect mesoscopic mechanical motion to atomic-scale quantum tunneling, establishing a platform where macroscopic forces can be sensed through what is in essence a quantum process.\\

The electronic set-up is straightforward. As in any STM configuration, only two electrodes are necessary; one to apply a bias voltage and one to read a current response. The current depends exponentially on the tip-sample distance, with typically a tenfold increase in current for every $0.15$~nm distance decrease, in our case. This extreme sensitivity allows for a highly responsive feedback loop to maintain a constant current, which can then be used to scan the surface topography. Here we use the same setup as with conventional STM, but with an oscillating sample. We introduce three complimentary measuring modalities and establish this platform's unique advantages for force measurements at cryogenic temperatures. Figure details such as applied power and tunneling setpoint are relegated to Supplementary Note 1. \\

\section{Experimental Set-up}
Our membrane samples consist of silicon substrates with deposited metal layers, which are patterned with holes to remove the underlying substrate, leaving behind suspended, high–aspect-ratio membranes. We focus on two types: a more compliant dielectric SiN membrane coated with superconducting NbTiN ~\cite{thoen2016superconducting} and a stiffer gold-coated membrane. We attribute the difference in compliance to the gold being deposited via e-beam evaporation, whereas the NbTiN is grown by atomic layer deposition (see \textit{Methods} section for more details). We find a Q-factor of over $10^6$ for the NbTiN coated membrane, see Supplementary Note 2. \\

To probe their delicate mechanical motion with atomic precision, we use an STM configuration that combines cryogenic operation, ultra-high vacuum ($\approx 10^{-11}$~mbar), and optical access that enables precise pre-alignment of the tip–sample junction on a patterned chip. The ultra-high vacuum environment minimizes viscous damping, typically difficult to achieve with standard turbo- or rough-pumped systems, thereby enhancing the resolution of the membranes’ intrinsic acoustic linewidths. Their high tensile stress keeps the membranes taut and stable, enabling reliable high-precision approaches. \\

\begin{figure*}[t]
    \includegraphics[width=\textwidth]{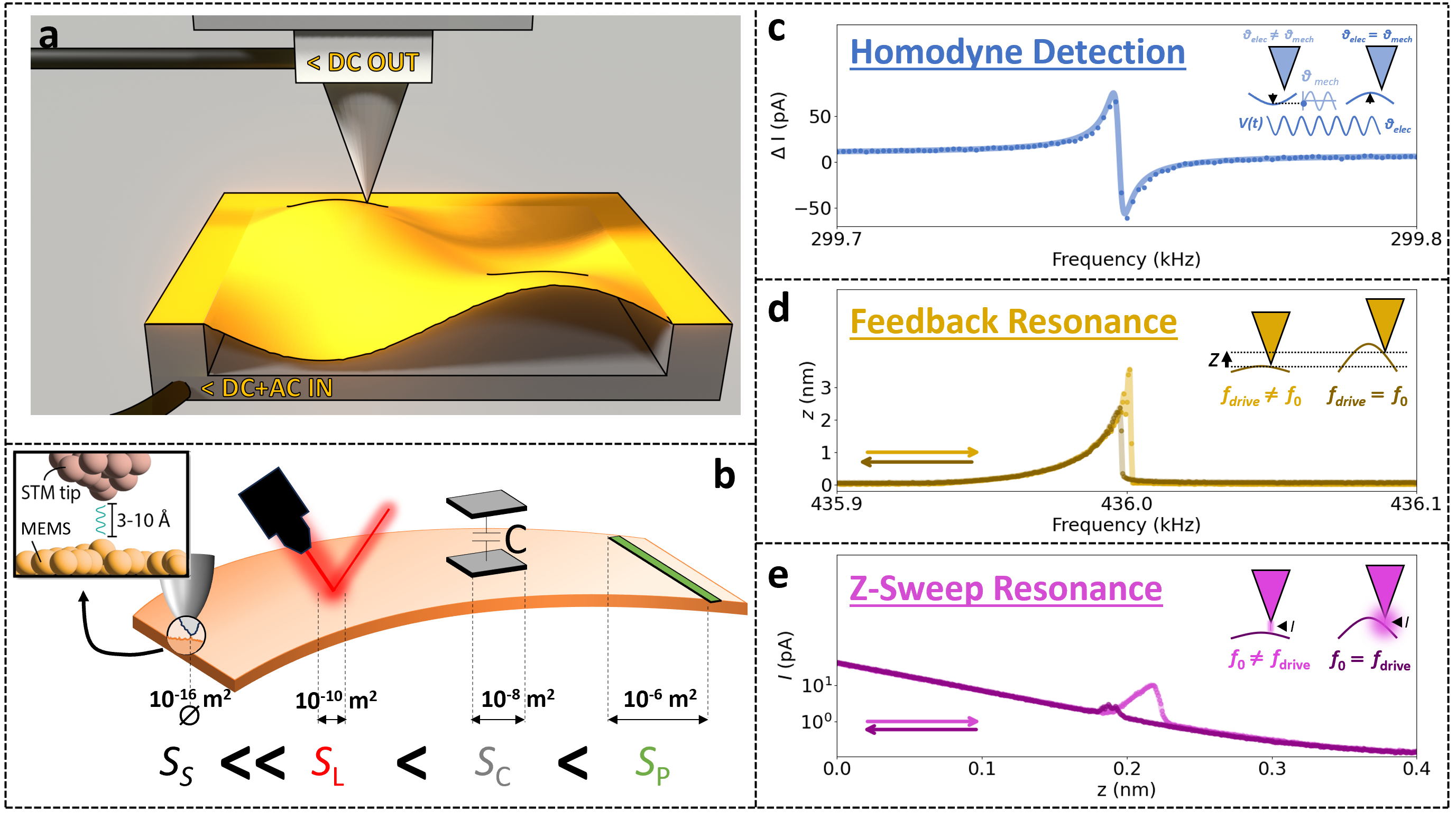}
    \caption{\textbf{Experimental set-up.} \textbf{(a)} A MEMS resonator containing a square superconducting membrane is supplied a DC and AC voltage, with a Scanning Tunneling Microscope (STM) tip probing the membrane. The oscillating voltage component drives the membrane motion, shown here in its (2,2) mode, while the tip measures a current as a result of this motion, which is ultimately converted to a DC current signal (see text). \textbf{(b)} Overview of interrogation windows of typical membrane motion detection techniques including: through an STM ($S_{\rm S}$), through optical means with a laser ($S_{\rm L}$), through capacitors ($S_{\rm C}$), and through piezoelectric signals ($S_{\rm P}$).   \textbf{(c-e)} Three different measurement modalities: Homodyne Detection \textbf{(c)} shows the lock-in signal, after adjusting the phase to put the capacitive background on a different channel (light curve is a Lorentzian fit), Feedback Resonance \textbf{(d)} shows the tip height and Z-sweep Resonance \textbf{(e)} shows current response. Forward and backward sweeps are shown by arrow colors.}
    \label{fig:1}
\end{figure*}

Figure \ref{fig:1}a outlines the experimental set-up. A combined DC + AC voltage is applied between the metallic membrane and the metal STM tip, following:
\[
V(t) = V_{\mathrm{DC}} + V_{\mathrm{drive}}\sin(2\pi f_{\mathrm{drive}} t),
\]

where the AC component with amplitude $V_{\mathrm{drive}}$ actuates the membrane motion while the DC offset $V_{\mathrm{DC}}$ stabilizes the tunneling current used by the STM feedback loop. The membrane oscillations are not observed in real time because the $1.1$~kHz bandwidth of the current pre-amplifier is far below the membrane's mechanical resonance frequencies, which in this study lie between $100$~kHz and $1$~MHz. Instead, we detect the motion indirectly through three different modalities, explained below. Using a nested feedback loop to maintain a stable tip current, we also maintain lateral precision down to the picometer scale. As shown later (Fig.~\ref{fig:xy-dependence}a,b), this positional stability is essential for reproducibility, since even subnanometer-scale lateral shifts in tip position alter the local tip–membrane coupling strength. This requirement arises from the STM’s exceptionally small effective interrogation area, set by the short-range nature of electron tunneling and near-field tip--sample forces, which is expected to be on the order of $S_{\rm S} \approx 10^{-16}$ m$^2$ (see Sec.~11.3 of Ref.~\cite{Israelachvili2011}). As illustrated in Fig.~\ref{fig:1}b, this nanoscale interrogation window distinguishes our approach from other membrane readout techniques such as optical interferometry (typical spot $S_{\rm L} \approx 10^{-10}$ m$^2$ \cite{Thompson2008Nature,Reinhardt2016PRX,norte2016}), capacitive sensing ($S_{\rm C} \approx 10^{-8}$ m$^2$ \cite{Faust2012PRL,andrews2014bidirectional,zawawi2020review}), or piezoelectric detection ($S_{\rm P} \approx 10^{-6}$ m$^2$ \cite{Mahboob2008APL,Elfrink2009JMM}). The STM thus accesses a spatially localized regime of membrane dynamics unattainable by macroscopic probe geometries. Furthermore, STM can provide unique insight into the electronic properties of the membrane (see Suppl. Note 3). \\

We explore three modalities of this measurement technique, illustrated in Fig.~\ref{fig:1}c-e: Homodyne Detection (c, blue), Feedback Resonance (d, yellow) and Z-sweep Resonance (e, pink). Each modality relies on aligning the driving frequency $f_{\rm drive}$ with the resonance frequency of the membrane $f_0$. In the first two modalities this is accomplished by sweeping $f_{\rm drive}$ while monitoring the response signal, whereas in the last modality it is $f_0$ that is indirectly swept under a constant $f_{\rm drive}$. \\

In the case of Homodyne Detection, the response corresponds to the difference in tunneling current between measurements with and without the oscillating voltage. This is achieved by chopping the excitation at a reference frequency $f_{\rm LI}$ well below the pre-amplifier cutoff and then demodulating the resulting signal at this same frequency. This yields a lock-in signal that carries phase information (see Supplementary Note 4 for more details), and is inspired by recent ESR-STM experiments, operating at GHz frequencies \cite{Hwang2022RevSciIns}. This lock-in signal scales approximately linearly with the driving amplitude and therefore imposes a lower limit to the usable drive power. Homodyne Detection, by virtue of multiplying an oscillating voltage with an oscillating conductance (through the oscillating tip–membrane distance) at equal frequency, produces a phase-sensitive DC signal scaling as $\cos({\vartheta_{\rm elec} - \vartheta_{\rm mech}})$, where $\vartheta_{\rm mech}$is the phase of the membrane's mechanical motion and $\vartheta_{\rm elec}$ is the phase of the oscillating electrical signal. This DC component is further enhanced by the exponential dependence of the tunneling conductance on the tip–membrane distance (see Supplementary Note 4). When combined with lock-in detection to reject out-of-band noise, this signal allows even sub-picometer membrane oscillations to be measured. Figure \ref{fig:1}c shows the lock-in signal, after rejecting the capacitive background via phase adjustment, measured on a stiff gold-coated membrane. \\

In more compliant membranes, such as those coated with NbTiN, the mechanical oscillation amplitudes can easily reach the nanometer scale and beyond. As the driving frequency $f_{\rm drive}$ approaches the membrane’s resonance frequency $f_0$, the membrane motional amplitude increases. This in turn reduces the effective tip-membrane distance, causing an increase in time-averaged tunneling current. The constant-current feedback loop then retracts the tip to maintain a stable time-averaged current, and we therefore take the tip height as the response signal in the Feedback Resonance modality. The abrupt drop shown in Fig.~\ref{fig:1}d occurs when $f_{\rm drive}$ exceeds $f_0$, and also reflects a softening response, as discussed later. When sweeping $f_{\rm drive}$ downward, a sudden jump appears instead. Provided that the feedback is sufficiently fast, a tip crash is avoided despite the sudden multi-nanometer oscillation amplitude. We note that the sudden drop shown during this modality also occurs in the membrane's linear regime (see Supplementary Note 5). \\

The Z-sweep Resonance modality inverts the logic of Feedback Resonance. Instead of sweeping the driving frequency, the tip height $z$ is varied while the feedback loop is disabled, and the tunneling current is recorded. As the tip is retracted and the tunneling current decreases exponentially, the forces imparted by the tip on the membrane decrease and the resonance frequency $f_0$ goes up. Then, when the membrane’s resonance frequency approaches the fixed driving frequency $f_{\rm drive}$, the membrane oscillation amplitude increases sharply and enhances the current. This produces a characteristic rise in current at a well-defined height $z_{\rm rise}$. Once the tip is moved beyond the maximum oscillation amplitude, the system abruptly switches to a lower-amplitude steady state, leading to a sudden drop in current. This behavior reflects the nonlinear softening response of the membrane, as explored later. 

In a reverse sweep (i.e., with decreasing tip height), only a minor current increase is observed, consistent with the hysteresis expected for a softening resonator. This modality can be applied with the tip positioned several nanometers above the surface which minimizes tip–membrane interaction. However, the absence of the feedback loop’s automatic retraction makes it more susceptible to tip crashes if operated at excessive driving power, and additional steps may be required for proper interpretation of results. \\

\UseTblrLibrary{booktabs}
\definecolor{z-sweep}{HTML}{D44CD1}
\definecolor{feedback}{HTML}{BA8600}
\definecolor{homodyne}{HTML}{4472C4}

\begin{table*}[t]
\centering
\footnotesize
\caption{Comparison of STM-based membrane resonance detection modalities, along three columns. Row number ($i$) describes: (1) key advantage of each modality, (2) transduced quantity, (3/4) whether the modality works with small and/or large mechanical oscillation amplitudes, (5/6) whether the modality is applicable with low and/or high driving power, (7) whether the modality requires the AC signal to be chopped and synchronized to a lock-in detector, (8) whether the STM feedback loop for a constant current is active during the modality, (9) whether the driving frequency is swept, or whether (indirect) the resonance frequency is swept and (10) an additional feature of each modality.}
\label{tab:modalities}
\begin{tblr}{
width=\linewidth,
  colspec = {X[0.3,l] X[1.5,l] X[1.8,j] X[1.8,j] X[1.8,j]},
  row{1} = {font=\bfseries},
  cell{1-Z}{1-Z} = {c,m}, 
  hlines,
  hline{1,2, Z} = {1.5pt},
  vlines = {1.5pt},
}
i & 
Parameter &
\textcolor{homodyne}{\textbf{Homodyne Detection}} &
\textcolor{feedback}{\textbf{Feedback Resonance}} &
\textcolor{z-sweep}{\textbf{Z-Sweep Resonance}} \\

1 &
Key advantage &
Provides phase information &
Quickest, safest modality &
Reaches non-perturbative limit \\

2 &
Signal &
$\Delta I$ (pA) &
$z$ (nm) &
$I$ (pA) \\

3 &
Works with SMALL amplitudes &
\checkmark  &
\textbf{---} &
\textbf{---}  \\

4 &
Works with LARGE amplitudes &
\checkmark &
\checkmark &
\checkmark \\

5 &
Applicable with LOW power &
\textbf{---} &
\checkmark &
\checkmark \\

6 &
Applicable with HIGH power &
\checkmark &
\checkmark &
\checkmark \\

7 &
Requires chopping &
\checkmark &
\textbf{---} &
\textbf{---} \\

8 &
Uses STM feedback &
\checkmark &
\checkmark &
\textbf{---} \\

9 &
Sweeps $f_{\rm drive}$ / $f_0$ &
$f_{\rm drive}$ &
$f_{\rm drive}$ &
$f_0$ \\

10 &
Additional feature &
Yields side peaks &
Can determine stiffness stability &
Allows determining long-distance force curve \\
\end{tblr}
\end{table*}

Table \ref{tab:modalities} summarizes the main differences in applicability between the three detection modalities. Homodyne Detection is preferred when phase information is required, for instance, to determine whether the oscillating electric field locally pushes or pulls the membrane, to verify that the electrical coupling between tip and membrane is stable, or to identify whether the response remains in the linear regime. Feedback Resonance is most suitable when measurements need to be performed rapidly, such as during limited cryogenic hold times, or when tracking the resonance frequency efficiently as it shifts under changing external conditions. For example, in our related work where we measure the Casimir force across the membrane's superconducting transition \cite{Xu2025arXiv}, we required low driving powers and a fast and reliable measurement modality to overcome our limited Liquid Helium hold time at elevated temperatures, thus making Feedback Resonance a natural choice. Z-Sweep Resonance, on the other hand, is the method of choice when one aims to minimize the influence of the STM tip on the membrane altogether and measure its intrinsic mechanical response in the least perturbative way. \\

Each modality operates within a characteristic range of oscillation amplitudes and driving powers. Because the Homodyne signal amplitude scales linearly with $V_{\rm drive}$, it is most effective at higher driving power. In contrast, the Feedback and Z-Sweep Resonance modalities require oscillation amplitudes exceeding a few picometers to produce measurable responses. Our gold-coated membrane did not yield high sufficient mechanical oscillation amplitudes, thereby limiting the applicable modality on this membrane to Homodyne Detection. 

Furthermore, Homodyne Detection requires the excitation signal to be chopped and demodulated at the lock-in frequency $f_{\rm LI}$ for acquiring an signal that overcomes the noise (see Supplementary Note 4 for more details). To achieve this sufficient signal-to-noise ratio, this method is slower than Feedback Resonance and slightly more technically involved. Interestingly, this same chopping introduces side peaks of the resonance at $f_i = f_{\rm drive} + (1+2i)f_{\rm LI}$, which may be used to help find resonance frequencies. Feedback Resonance, in contrast, can reveal stiffness instabilities by identifying conditions under which the tip transitions between two metastable states (see Supplementary Notes 5 and 7). Finally, Z-Sweep Resonance uniquely probes the non-perturbative limit and can map the full force curve as the tip retracts, as illustrated in Fig.~\ref{fig:z-dependence}. \\

In the following sections, we first describe the main forces at play and how this would affect, for example, the Z-Sweep Resonance results. Along the way we show that both the Z-sweep Resonance and the Feedback Resonance can be used to determine frequency shifts as a result of the electrostatic force by changing the DC bias voltage $V_{\rm DC}$. We will then show the lateral spatial dependence of this overall technique, both at the microscopic and the macroscopic scale. Afterwards, we explore the entire force curve through the Z-sweep resonance and determine the force resolution in the non-perturbative limit. \\

\section{Forces}

\begin{figure}[h]
    \centering
    \includegraphics[width=\linewidth]{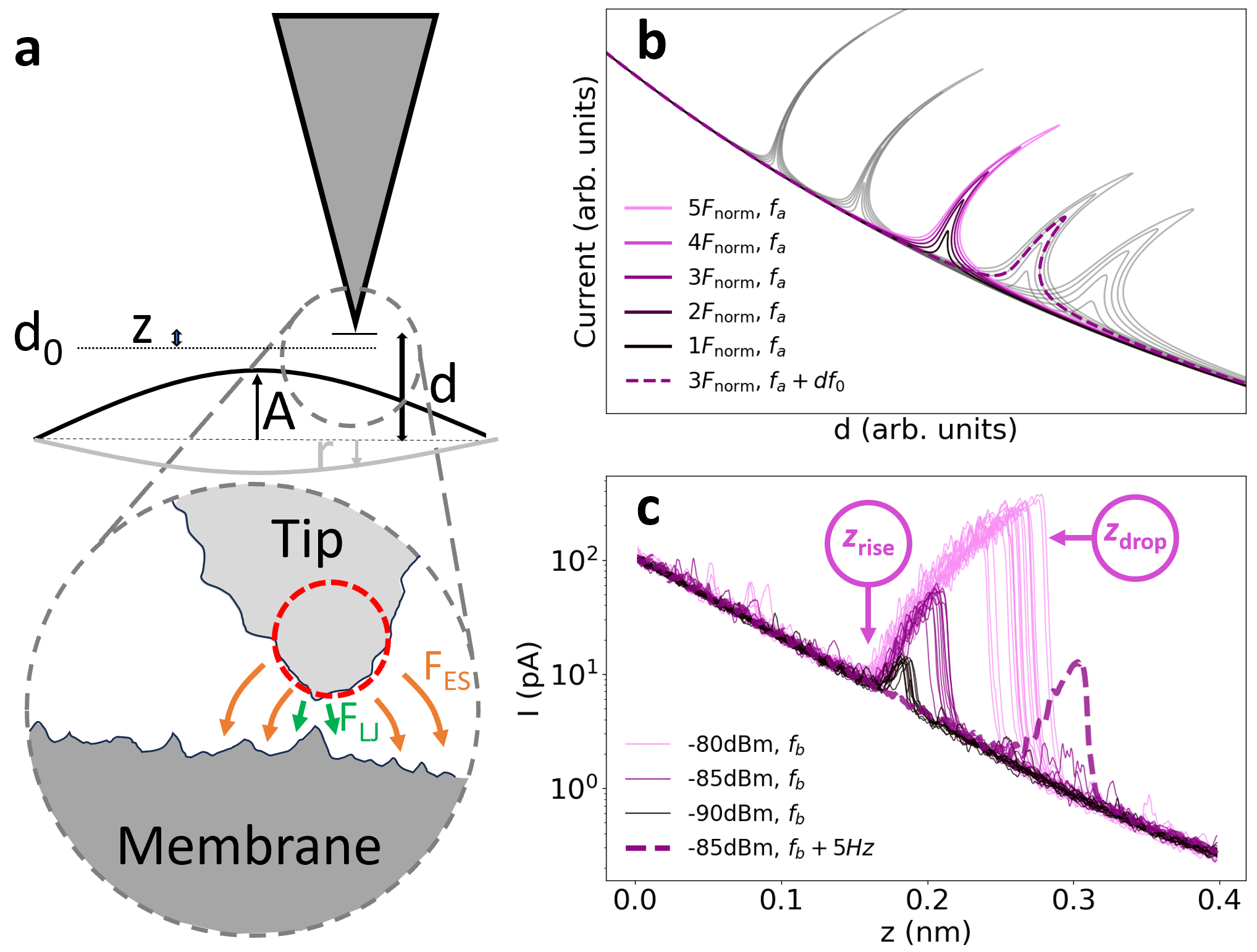}
    \caption{\textbf{Force Model} \textbf{(a)} Diagram of relevant forces (Lennard-Jones force $F_{\rm LJ}$ and electrostatic force $F_{\rm ES}$) and the definition of amplitude $A$ and tip-membrane distance $d$, an arbitrary offset $d_0$, the distance to this offset $z$ and the membrane position variable $r$. The red circle indicates spherical approximation of the tip. \textbf{(b)} Simulated currents as described in the \textit{Methods}. Five different driving frequencies are simulated, resulting in canted peaks in the current at five different tip heights $d$. Each set is repeated five times with different driving powers $P$. Legend shows relative driving power (see text). Six simulation results are highlighted in different colors. \textbf{(c)} Experimental data on a NbTiN membrane, with multiple measurements taken at a driving frequency of $278.61$~kHz for various powers, and one taken at $278.615$~kHz, $-85$~dBm. Shown are also $z_{\rm rise}$ and $z_{\rm drop}$.}
    \label{fig:concept}
\end{figure}

We now examine the fundamental forces between the STM tip and the membrane, predominantly the Lennard-Jones force $F_{\rm LJ}$ and the force derived from the electrostatic potential $F_{\rm ES}$ (Fig.~\ref{fig:concept}a). For a full mathematical description,  please see the \textit{Methods} section. Here we will focus on a more qualitative description.

In the linear regime, the unperturbed membrane behaves as a harmonic oscillator of stiffness $k$, while the tip introduces an additional nonlinear potential. The combined potential $U(r)$ and its gradient yield an effective force $F(d,V_{\rm DC})$, whose curvature shifts the resonance frequency. As shown in the \textit{Methods} section, the effective force is given as:

\begin{equation}\label{eq:forces_1}
    F(d,V_{\rm DC}) =  4 \epsilon_{\rm LJ}  \left[-12 \left( \frac{\sigma^{12}}{d^{13}} \right) +6 \left( \frac{\sigma^{6}}{d^{7}} \right) \right] - \frac{C_1 V_{\rm DC}^2}{d^p}
\end{equation}

while the associated frequency shift can be expressed as:

\begin{equation}\label{eq:freq_shift}
     \Delta f^2 = \frac{1}{4\pi^2} \left[ \frac{4\epsilon_{\text{LJ}}}{m} \left( \frac{156\sigma^{12}}{d^{14}} - \frac{42\sigma^6}{d^8} \right) -\frac{pC_1V_{\rm DC}^2}{m d^{p+1}} \right].
\end{equation}

\noindent Here $\epsilon_{\text{LJ}}$ is the energy coefficient, $m$ is the modal mass, $\sigma$ is the distance parameter, $d$ is the distance between the tip and the equilibrium position of the membrane and $C_1$ is the effective electrostatic coupling constant between tip and membrane and $1 \leq p \leq 2$ depends on the exact tip shape (i.e. to what extend the red circle approximates the tip well in Fig.~\ref{fig:concept}a). Furthermore, $d$ is the actual tip-membrane distance, while $z$ is a relative measure from an initial starting point $d_0$, based on an initial conductance: $d = z + d_0$. \\

We will show how this model applies to the results obtained with the Z-Sweep Resonance, as this modality uniquely offers  a view into the entire force curve. Numerical simulations (see Fig.~\ref{fig:concept}b) using this model reproduce the nonlinear response observed experimentally (see Fig.~\ref{fig:concept}c) with the Z-sweep Resonance modality. Here we increase the drive power from $F_{\rm norm}$ to $5\cdot F_{\rm norm}$, where $F_{\rm norm}$ is a normalized force qualitatively determined to be at the onset of nonlinear frequency response. The increased drive power amplifies nonlinear response, producing canted peaks characteristic of nonlinear softening (see \textit{Methods} section for more details). The experimental data in Fig.~\ref{fig:concept}c show corresponding behavior, including a sudden drop in current as the system transitions to a lower-amplitude steady state. We note that occasional switching between these states (see Supplementary Note 6) confirms the presence of bistability typical of softening resonators. \\

For these softened peaks, accurately determining the exact value of $z_0$, the tip height where the resonance frequency precisely matches the driving frequency, can be challenging. However, as shown in Fig.~\ref{fig:concept}c, since $z_{\rm rise}$, the tip height during tip retraction where the current starts to increase, is relatively stable across driving power while $z_{\rm drop}$, the tip height during tip retraction where the current suddenly drops, is less consistent, we approximate $z_0 \approx z_{\rm rise}$. To determine the actual resonance frequency $f_0$, especially as it varies with other parameters, a more sophisticated analysis is necessary, as illustrated in the next section.

\begin{figure}[h]
    \centering
    \includegraphics[width=\linewidth]{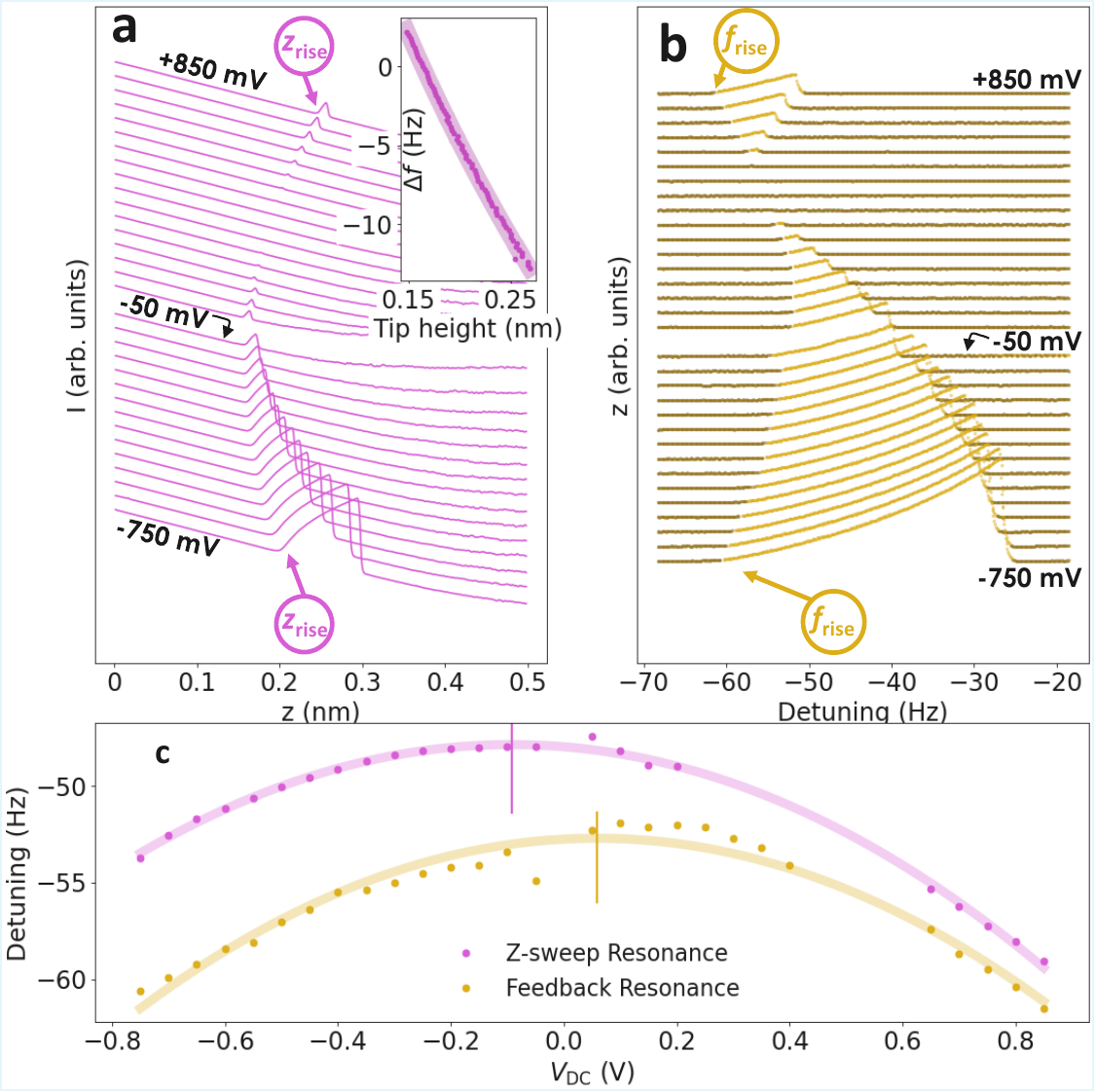}
    \caption{\textbf{Local Contact Potential Differential} \textbf{(a)} Z-sweep Resonance curves, showing the current as a function of tip height $z$, for various bias voltages $V_{\rm DC}$. All curve starting points (i.e. $z = 0$~nm) are offset according to the applied $V_{\rm DC}$. The tip height where the current starts to increase is shown as $z_{\rm rise}$. Inset: mapping of tip height to frequency shift (see text for more details) to turn $z_{\rm rise}$ into $f_{\rm rise}$. \textbf{(b)} Feedback resonance curves, showing the tip height as a function of drive frequency $f_{\rm drive}$, for various bias voltages $V_{\rm DC}$. All curve starting points (i.e. detuning = -70Hz) are offset according to the applied $V_{\rm DC}$. The frequency of increasing tip height is shown as $f_{\rm rise}$. \textbf{(c)} Detuning of resonance frequency as derived from panels a and b and a parabola fit for both data sets, yielding a local contact potential difference of $-92.6 \pm 4.3$~mV using the Z-sweep Resonance measuring modality and $+58.3 \pm 10.4$~mV using the Feedback Resonance measuring modality. Reference frequency $f_0^*$ to determine the detuning is derived from the fit shown in Fig.~\ref{fig:z-dependence}a.}
    \label{fig:lcpd}
\end{figure}

\section{Electrostatic Forces}
The last term in Eq.~ \ref{eq:freq_shift} is determined by the electrostatic force and scales quadratically with $V_{\rm DC}$. Varying this potential allows extraction of the local contact potential differential (LCPD) $V_{\rm LCPD}$, using either the z-sweep and feedback resonance modalities. This approach is analogous to Kelvin-probe measurements in AFM \cite{Bettac2009Nanotech}, where $V_{\rm DC}$ compensates the chemical potential offset between tip and membrane. Figure \ref{fig:lcpd}a shows z-sweep resonance curves at different $V_{\rm DC}$. As $|V_{\rm DC}|$ increases, the onset of current rise $z_{\rm rise}$ shifts to larger tip-membrane distances, indating that the electrostatic attraction modifies the resonance frequency $f_0$. Converting these $z_{\rm rise}$ using a height-to-frequency mapping (inset) yields a parabolic dependence of frequency detuning on bias (panel c), characteristic of LCPD behavior. The mapping in the inset was made by finding $z_{\rm rise}$ in repeated Z-Sweep Resonance measurements at different $f_{\rm drive}$. \\

A similar trend appears for the feedback resonance data in Fig.~\ref{fig:lcpd}b, where the resonance frequency $f_0$ is estimated with the onset of the tip retraction  $f_{\rm rise}$ during upward frequency sweeps. Both modalities reproduce the expected LCPD parabolas (panel c), with Z-sweep resonance  offering higher precision (uncertainty is less than half) but requiring additional acquisition steps. The detuning shown in the axes of panels b and c are with respect to $f_0^*$, which corresponds to the resonance in absence of the tip, discussed later. We note that, as shown in Supplementary Note 8, when the tip is brought closer (i.e. at a larger setpoint current) the LCPD parabola gets steeper. \\

\section{Lateral dependence}

Having established the electrostatic contribution to the tip–membrane interaction, we next examine how this coupling varies spatially across the membrane. These measurements serve two purposes: first, they demonstrate the exceptional positional reproducibility of the STM-based approach, allowing us to return to the same nanoscopic position, even to the level of individual surface atoms. Second, they allow us to assess the uniformity of the mechanical and electrical response across both microscopic and macroscopic length scales. This spatial validation provides the foundation for the subsequent exploration of the full interaction potential towards the non-perturbative limit. \\

\begin{figure}[t]
    \centering
    \includegraphics[width=\linewidth]{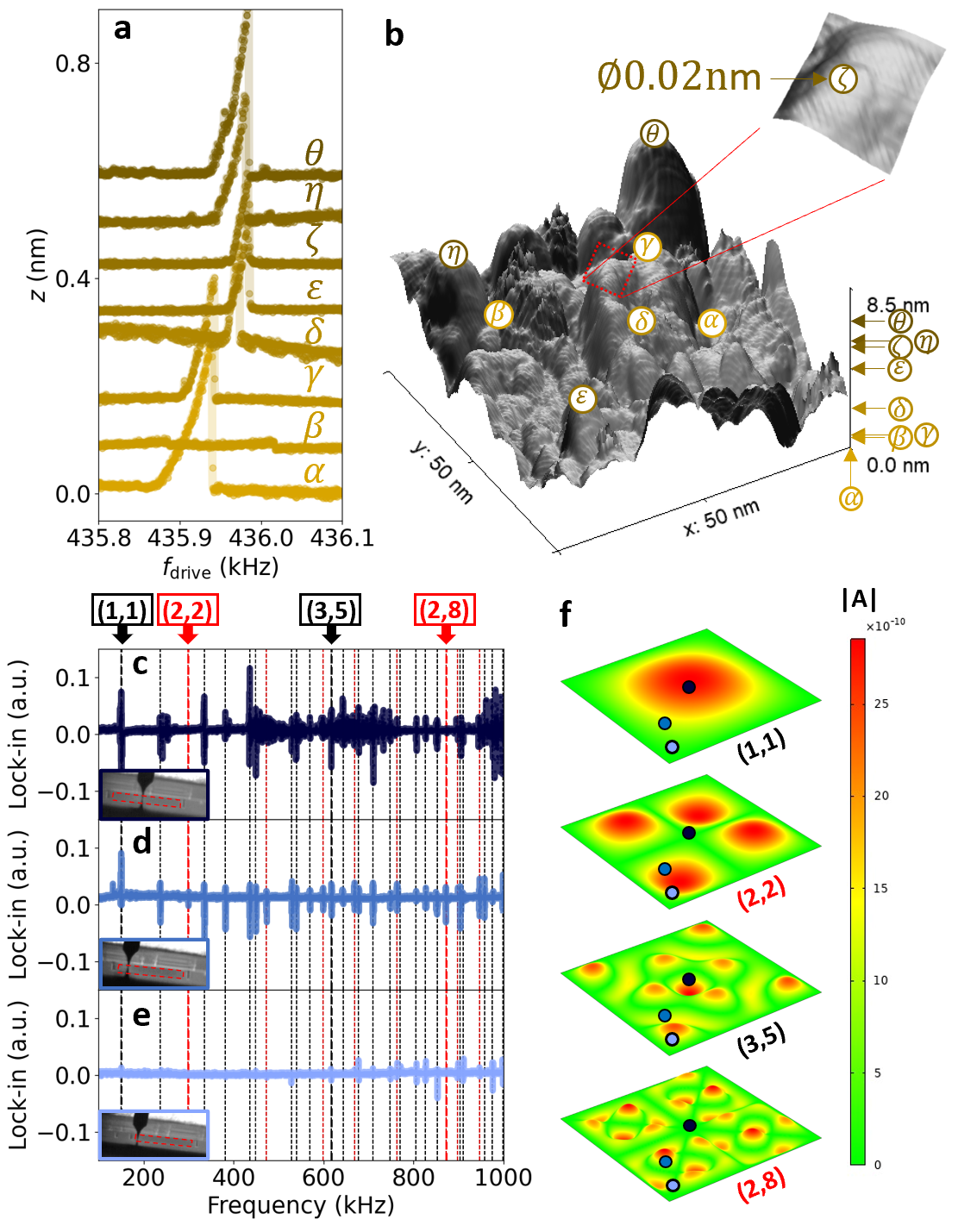}
  \caption{\textbf{Lateral dependence}  \textbf{(a)} Feedback Resonance measurements taken at various locations shown in panel \textbf{b} with corresponding Greek letters. Each curve is offset by for clarity. \textbf{(b)} Topographic image on which the data of panel \textbf{a} is taken, shown in three dimensions. Height axis indicates the off-resonance height of each measurement. Zoom-in indicates the accuracy with which the tip is able to maintain the same position to be $0.02$~nm. \textbf{(c)} Homodyne Detection measurements taken between $100$~kHz and $1000$~kHz on the gold-coated membrane. With the (1,1) mode at $150$~kHz, the vertical lines indicate the expected frequencies of higher modes. Red lines indicate modes not observed in panel c. Inset shows photograph of tip location. \textbf{(d,e)} Same as d, but for a different top positions. \textbf{(f)}  Finite elements simulations of four selected modes on a square membrane. Colors indicate the absolute value of the amplitude. Three dots per membrane correspond to tip locations of panels \textbf{c}-\textbf{e} with corresponding colors. }\label{fig:xy-dependence}
\end{figure}

The resonance frequencies of the membrane vary laterally on the microscopic scale, primarily because the STM tip samples slightly different effective tip--membrane distances as it moves across the surface topography at constant current. Figure~\ref{fig:xy-dependence}a shows Feedback Resonance measurements performed at several positions in close proximity to each other on a NbTiN membrane (locations marked in panel~b). Even lateral shifts of only a few nanometers produce distinct resonance frequencies and amplitudes. This sensitivity reflects how local topographic variations lead to measurable changes in the tip–membrane coupling. We found, in general, an increase in resonance frequency when the tip is positioned at higher absolute position, However,  this dependence is not strictly monotonic due to the irregular and unknown shape of the tip (i.e. Fig.~\ref{fig:concept}a zoom-in is highly idealized). Likewise, a stronger response is typically found at local height maxima. Both observations are consistent with the tip being further away from the bulk of the membrane. To reach these local maxima, we employ nested “top-of-hill’’ feedback routines that keep the tip centered on such height maximum, allowing positional stability on the order of $\approx0.02$~nm (see Fig.~\ref{fig:xy-dependence}b inset). Maintaining the membrane interrogation position with such precision is one of the key aspects that sets this STM-based method apart. \\

On larger length scales, the position dependence reflects the membrane’s vibrational mode structure rather than local topography. To explore this regime, we employ Homodyne Detection on a gold-coated membrane, whose low compliance suppresses large oscillation amplitudes. This allows us to apply strong driving forces and perform stable, wide-range frequency sweeps without risking tip–membrane contact. Figures~\ref{fig:xy-dependence}c--e display representative lock-in responses obtained with the STM tip positioned at the membrane center, at an intermediate location, and near a corner (see insets). The spectra span 100~kHz--1000~kHz and show multiple resonances corresponding to higher-order mechanical modes.

Using the fundamental (1,1) mode at \( f_{1,1} = 150~\text{kHz} \) as a reference, the expected frequencies of higher modes are
\[
f_{i,j} = f_{1,1}\,\frac{\sqrt{i^{2} + j^{2}}}{2}.
\]
Vertical dashed lines in the spectra mark these predicted mode positions. For each mode index \((i,j)\), the STM detects a peak only when the tip sits at a position where that mode exhibits appreciable displacement (i.e., away from a node). This explains, for example, why modes with center nodes (e.g., \((2,2)\) or \((2,8)\)) vanish when the tip is at the center, and why corner placement suppresses modes whose displacement is concentrated toward the membrane interior. Finite-element simulations (COMSOL~Multiphysics) in Fig.~\ref{fig:z-dependence}f confirm this spatial selectivity, implying we can position our tip such to access specific modes only.

These results show that STM-based actuation and readout provide a localized probe of membrane motion: small lateral shifts change the effective tip--membrane coupling, and deliberate repositioning maps the modal landscape.

\section{Long-range sensitivity and force resolution}
Having shown that small lateral displacements alter the local coupling strength, we now examine how the same coupling evolves with vertical displacement. Here, we capture the transition from the perturbative domain, where the tip–membrane interaction noticeably shifts the resonance frequency, to the non-perturbative limit, where the membrane remains driven but oscillates at its intrinsic resonance with only minute frequency shifts. We will then translate those into measurable force differences. \\

\begin{figure}[t]
    \centering
    \includegraphics[width=\linewidth]{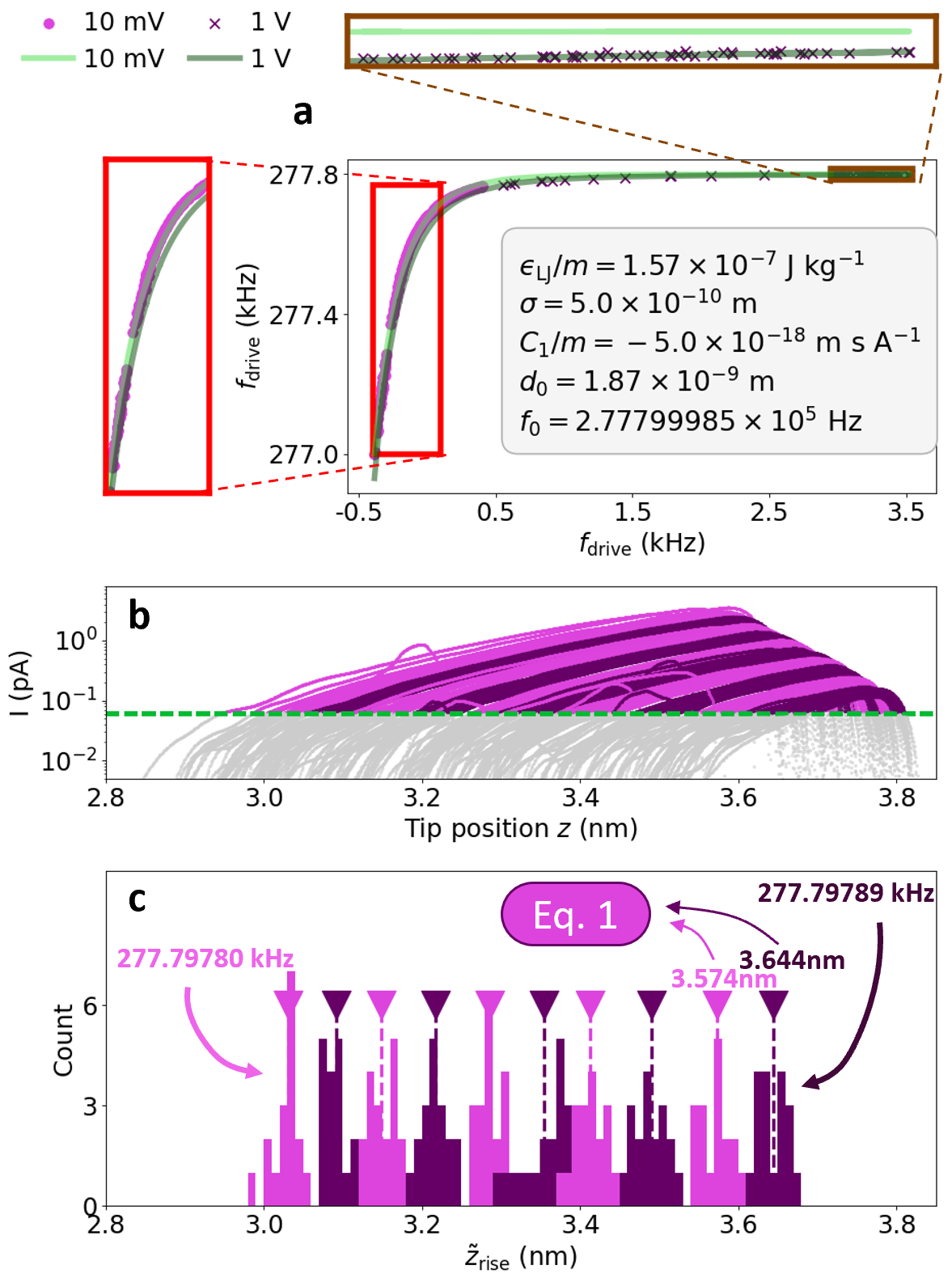}
    \caption{\textbf{Long-range frequency sensitivity and force resolution.} \textbf{(a)} Resonance frequency versus tip height $z_{\rm rise}$ over a 4~nm range, measured at $V_{\rm DC}=10$~mV (circles) and $V_{\rm DC}=1$~V (crosses). Separate fits to Eq.~\ref{eq:freq_shift} are shown for each bias. Insets highlight the regimes where each fit works well. \textbf{(b)} Twenty repeats of 10 different frequencies, stepped in 10~mHz. The green dashed line indicates the threshold from which $\tilde{z}_{\rm rise}$ is determined. \textbf{(c)} Histograms of $\tilde{z}_{\rm rise}$ from data shown in panel b. The lowest and highest frequencies are shown. Mean of histograms areindicated by triangles. Using the values related to the two highest frequencies and Eq.~\ref{eq:forces_1} we determine a detectable force difference of $\approx$6~pN.}
    \label{fig:z-dependence}
\end{figure}

The long-range dependence of the resonance frequency on tip--membrane separation is shown in Fig.~\ref{fig:z-dependence}a, covering a displacement range of nearly 4~nm, acquired using the Z-sweep Resonance modality. At large separations, the membrane approaches its unperturbed resonance frequency $f_0^*$ and the curve asymptotically flattens. This behavior reflects the decay of both the Lennard--Jones and electrostatic forces with distance, as described by Eq.~\ref{eq:forces_1}. In Fig.~\ref{fig:z-dependence}a two sets of data appear: measurements taken at $V_{\rm DC}=10~\text{mV}$ (circles) and at $V_{\rm DC}=1~\text{V}$ (crosses). The low bias data are used for small tip–membrane separations to avoid preamplifier saturation, whereas the higher bias is required at larger separations to maintain sufficient signal-to-noise. Because the electrostatic force scales as $V_{\rm DC}^2$, the two bias conditions produce slightly different frequency–distance curves. For clarity, separate fits to Eq.~\ref{eq:freq_shift} are shown for each bias (light and dark green curves), using identical Lennard–Jones parameters (shown in the Figure) but different values for $V_{\rm DC}$. Both fits reproduce their respective data ranges well and together describe the full long-range behavior. \\

We note that the fit returns an apparent offset $d_0 = 1.87\,\mathrm{nm}$, which exceeds typical static STM tunneling gaps. We interpret this offset as arising from time-averaging over fast, thermally populated non-resonant membrane motion, due to the low bandwidth of the current readout ($\leq 1.1\,\mathrm{kHz}$) compared to the membrane mode frequencies ($\gtrsim 100\,\mathrm{kHz}$). This contrasts with our electrically driven resonant measurements; however, in both cases the majority of the tunneling current flows near the point of maximum membrane extension (see Supplementary Note~9 for details).\\ 

In the non-perturbative limit (i.e. large $z$), the frequency–distance curve approaches an asymptote, and its slope $df/dz$ becomes very small. A small shift in resonance frequency therefore maps onto a large change in effective tip height. As a result, $z_{\rm rise}$ would be an extremely sensitive proxy for minute variations in force. However, the exact position of $z_{\rm rise}$ cannot be determined as this would occur well below the noise floor. Instead, we opt for the quantity $\tilde{z}_{\rm rise}$, which we define as the $z$ value where the current has roughly exceeded $60$~fA. This threshold is shown as the green dashed line in Fig.~\ref{fig:z-dependence}b, see Supplementary Note 10 for more details. To quantify the sensitivity, Fig.~\ref{fig:z-dependence}b shows a representative zoom of the high-$z$ tail ($z_{\rm rise} \gtrsim 2.8$~nm), corresponding to the rightmost portion of Fig.~\ref{fig:z-dependence}a where the force curve goes nearly horizontal. For each drive frequency (stepped in 10~mHz increments), the figure shows twenty repeated measurements. The resulting distribution of $\tilde{z}_{\rm rise}$ values, shown in Fig.~\ref{fig:z-dependence}c, reveals narrow histograms demonstrating that frequency differences as small as 10~mHz can be robustly resolved.

This frequency precision can be converted into a force sensitivity by evaluating the force difference between two closely spaced tip--membrane distances. Using Eq.~\ref{eq:forces_1} and the two $\tilde{z}_{\rm rise}$clusters in Fig.~\ref{fig:z-dependence}b that correspond to the largest tip separation, (3.574~nm and 3.644~nm), we obtain a measurable differential force of approximately 6~pN (see Supplementary Note~11 for details). 

While the present implementation already yields piconewton-level sensitivity, further improvements are achievable. Operating closer to the non-perturbative limit (lower \(V_{\rm DC}\)), increasing integration time, or employing a higher-gain current preamplifier would reduce the effective noise floor. For example, repeating these measurements at 10~mV bias, rather than 1~V, which yields a stronger electrostatic force gradient, would lower the calculated force resolution to roughly 50~fN. Higher-Q modes or cryogenic optimization could reduce this figure further.

Together, the long-range frequency map and its asymptotic tail demonstrate that STM-based membrane actuation, when operated in the non-perturbative limit, enables extremely sensitive detection of tip-induced forces, which may be used to quantify external forces.

\section{Conclusions}
We have demonstrated how a low-temperature STM tip can probe the dynamics of an oscillating membrane by exploiting the electric forces between the tip and membrane that drive and sense mechanical resonance modes. These forces can be precisely tuned to match the membrane’s natural frequencies with an external generator, enabling stable and minimally invasive measurements across several operating modalities. By integrating this approach with high-stress nanomechanical membranes that offer exceptional stability, we achieve sub-nanometer positional control on on-chip mechanical structures. The combined use of Z-Sweep Resonance, Feedback Resonance, and Homodyne Detection provides complementary information, offering extensive insight into the membrane dynamics. Together, these capabilities open a practical pathway for studying mesoscopic membrane acoustics and, more broadly, the interplay between mechanical and electronic properties in cryogenic and magnetic-field environments, all while minimizing damping and external interference.

Future work could enhance frequency resolution further, while the method’s sensitivity to large-scale forces, such as acceleration, pressure, and surface forces, positions it as a promising tool in quantum sensing applications. Examples include detecting Casimir forces \cite{Xu2025arXiv} or monitoring nanoscale growth in real time. Furthermore, by combining this technique with higher Q-factor membranes, it may be possible to explore more complex macroscopic quantum phenomena, making STM-based nanomechanical sensing a versatile platform for both fundamental research and applied science.

\newpage
\section{Methods}

\begin{figure}[t]
    \includegraphics[width=\linewidth]{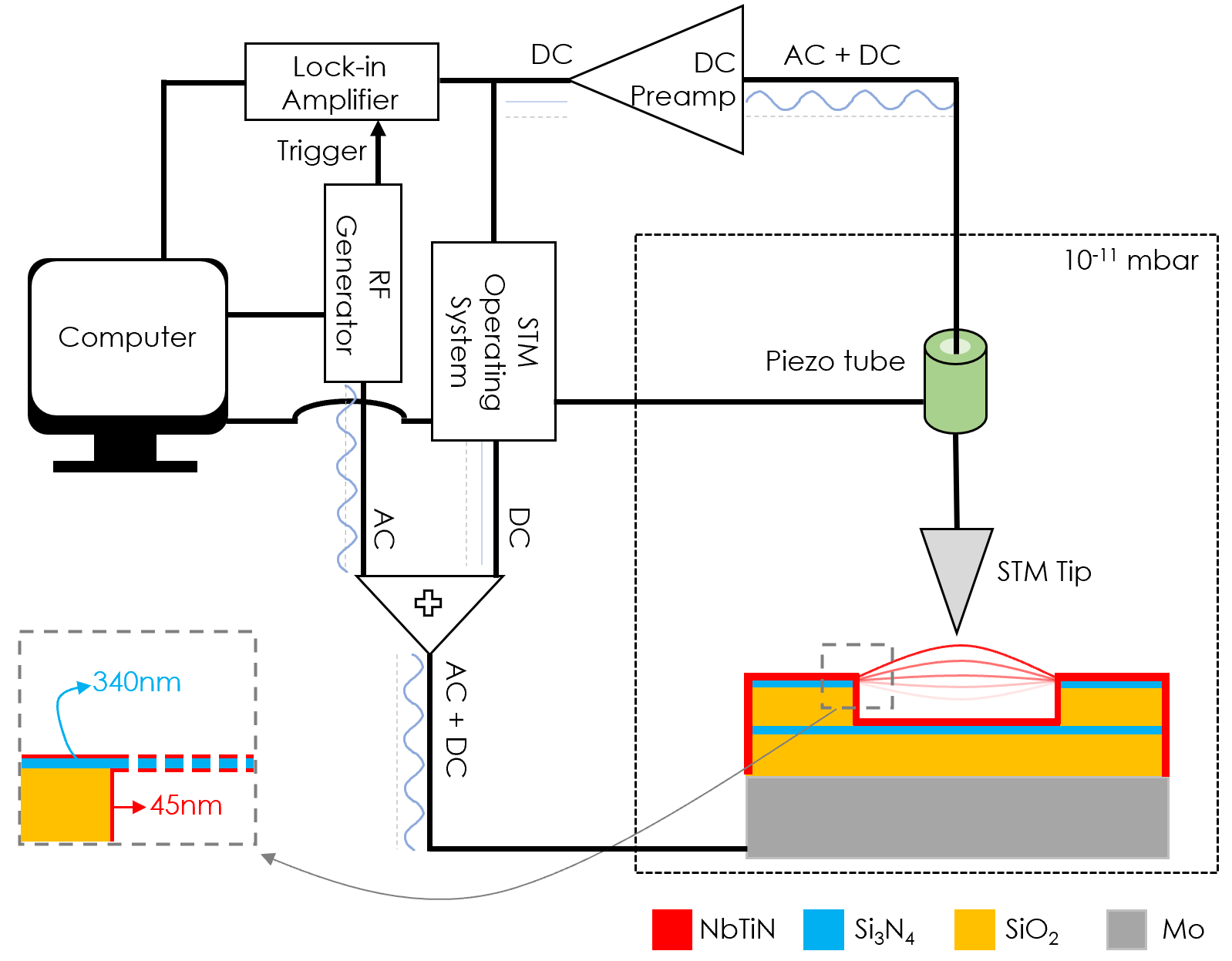}
    \caption{\textbf{Electronic setup.} \textbf{(a)} Schematic of the STM-based actuation and readout system. A combined DC and AC voltage is applied to the chip containing the membrane, while a Scanning Tunneling Microscope (STM) tip measures the resulting tunneling current. The AC excitation drives membrane motion, and the low-pass-filtered current is used either for feedback control. See text for more details.}
    \label{fig:electronic}
\end{figure}

\subsection{Electronic Setup}
For this study, we use a Joule--Thomson Scanning Tunneling Microscope (JT--STM) by SPECS\textsuperscript{\texttrademark}, equipped with optical access. The STM scanner can be cooled to $\sim 1$~K through a combination of liquid nitrogen, liquid helium, and a Joule--Thomson cooling cycle. The STM chamber is pumped to about $10^{-10}$~mbar, and the shielded volume around the scanning probe reaches pressures of $10^{-11}$~mbar or below. \\

Figure~\ref{fig:electronic} shows the electronic setup. A tungsten STM tip is mounted on a piezo tube controlled by the STM operating system (Nanonis\textsuperscript{\texttrademark}). The piezo tube adjusts the tip height dynamically, either via a constant-current feedback loop or, in the case of the Z-Sweep Resonance modality, through controlled open-loop sweeps. In feedback mode, the measured current is continuously compared to a setpoint (target) current. If the measured current falls below the setpoint, the operating system increases the piezo voltage, moving the tip closer to the membrane, and vice versa. \\

The tunneling current is low-pass filtered by a FEMTO DLPCA-200 preamplifier with a bandwidth of $1.1$~kHz, providing a transimpedance gain of $10^{9}$~V/A. This bandwidth limits both the response speed of the feedback loop and our ability to resolve the membrane’s true mechanical motion in real time, but allows us to measure currents with pA precision.  \\

The STM electronics apply a steady DC bias voltage, which maintains the tunneling current and can be used to adjust the chemical potential (see Fig.~\ref{fig:lcpd}). An RF generator (Rohde \& Schwarz SMB100A) provides an AC excitation in the range of $100$~kHz to $1$~MHz. The DC and AC signals are combined with a bias tee (Mini-Circuits ZFBT-4R2GW+) and routed to the sample holder through the JT--STM wiring. \\

The sample holder is made of molybdenum, with tantalum clips securing the experimental chip via molybdenum screws. Because the chip includes two insulating layers (illustrated in blue in Fig.~\ref{fig:electronic}), the electrical signal is routed only to the top surface through the tantalum clips, ensuring that the AC and DC voltages couple to the membrane. Figure~\ref{fig:electronic} illustrates the configuration for the NbTiN-coated membrane; an otherwise identical gold-coated chip is described in the subsection \textit{Sample Preparation}. \\

Throughout the experiment, the AC signal was chopped using the internal modulation feature of the RF generator, with a duty cycle of $50\%$ and a modulation frequency $f_{\rm LI} \leq 1.1$~kHz. This chopping effectively reduces the time-averaged applied power by 3~dB. The RF generator also provides a trigger reference to a lock-in amplifier (Stanford Research SR830), which demodulates the low-pass-filtered tunnel current and thereby measures the difference between RF-on and RF-off phases. We kept the chopping enabled even in measurements where no Homodyne Detection signal was recorded, as it did not adversely affect the other modalities. One example of a combined dataset between Homodyne Detection and Feedback Resonance is shown in Supplementary Note 5. \\

\subsection{Mathematical Model}
Here we expand upon the mathematical description provided in the main text. We consider the fundamental mode of the membrane with the generalized coordinate $r$, which measures the membrane displacement relative to its equilibrium position. In additional to the instrinsic harmonic restoring force, the membrane experiences an interaction with the STM tip, dominated by Lennard-Jones and electrostatic contributions, (see Fig.~\ref{fig:concept}a). In the absence of tip-sample interactions, the membrane behaves as a harmonic oscillator of stiffness $k$. Including the tip-induced interactions, the total potential energy of the system can be expressed as:

\begin{equation}\label{eq:potential}
    U(r) = \frac{1}{2}kr^2 + 4 \epsilon_{\rm LJ}  \left[ \left( \frac{\sigma}{r + d} \right)^{12} - \left( \frac{\sigma}{r + d} \right)^{6} \right] - \frac{C_1 V_{\rm DC}^2}{r+d}
\end{equation}

\noindent Here the electrostatic potential is  From this we can derive the force $F(r) = - \frac{\partial U}{\partial r}$ imparted by the tip on a static membrane as:

\begin{equation}\label{eq:forces}
    F(d,V_{\rm DC}) =  4 \epsilon_{\rm LJ}  \left[-12 \left( \frac{\sigma^{12}}{d^{13}} \right) + 6 \left( \frac{\sigma^{6}}{d^{7}} \right) \right] - \frac{C_1 V_{\rm DC}^2}{d^2}
\end{equation}

\noindent Assuming modal mass m and kinetic energy $T=\frac{1}{2}m\dot{r}^2$, we can express the Lagrangian as $\mathcal{L} = T - U$, which results in the following Lagrange equation of motion, $\frac{d}{dt} \left( \frac{\partial \mathcal{L}}{\partial \dot{r}} \right) - \frac{\partial \mathcal{L}}{\partial r} = 0$:

\begin{equation}
\begin{aligned}
m\ddot{r} + k r & + 4\epsilon_{LJ} \left( -12 \frac{\sigma^{12}}{(r + d)^{13}} + 6 \frac{\sigma^6}{(r + d)^7} \right) \\
& + \frac{C_1 V_c^2}{(r + d)^2} = 0
\end{aligned}
\end{equation}

The effective resonance frequency of the mode can be approximated by taking the first-order Taylor expansion of the forces around $r=0$, resulting in the effective linear stiffness $k_{\rm eff}$ and effective resonance frequency $\omega_{\rm eff}$:
\begin{equation}
k_{\rm eff}=k + 4 \epsilon_{\text{LJ}} \left( \frac{156\sigma^{12}}{d_{0}^{14}} - \frac{42\sigma^6}{d^8} \right) -\frac{2C_1V_c^2}{d^3}
\end{equation}
\begin{equation}
 \omega_{\rm eff}=\sqrt{\frac{k_{\rm eff}}{m}}=\sqrt{\omega_0^2 + \frac{4\epsilon_{\text{LJ}}}{m} \left( \frac{156\sigma^{12}}{d^{14}} - \frac{42\sigma^6}{d^8} \right) -\frac{2C_1V_c^2}{m d^3}}
\end{equation}
where $\omega_0$ is the linear resonance frequency. The shift of the resonance frequency squared can be conveniently expressed as:
\begin{equation}
 \Delta\omega^2=\omega_{\rm eff}^2-\omega_0^2= \frac{4\epsilon_{\text{LJ}}}{m} \left( \frac{156\sigma^{12}}{d^{14}} - \frac{42\sigma^6}{d^8} \right) -\frac{2C_1V_c^2}{m d^3}
\end{equation}

Thus

\begin{equation}\label{eq:freq_shift_2}
     \Delta f^2 = \frac{1}{4\pi^2} \left[ \frac{4\epsilon_{\text{LJ}}}{m} \left( \frac{156\sigma^{12}}{d^{14}} - \frac{42\sigma^6}{d^8} \right) -\frac{2C_1V_{\rm DC}^2}{m d^3} \right]
\end{equation}.

This expression corresponds to the special case $p=2$ of the phenomenological electrostatic force introduced in the main text. During the fit of Fig.~\ref{fig:z-dependence}a, a generalized power-law dependence $F_{\rm ES}\propto V_{\rm DC}^2/d^p$ is employed, and the best agreement with the measured frequency shifts is obtained for $p\simeq 2$, corresponding to a force-gradient scaling $\partial_d F_{\rm ES}\propto d^{-3}$. This corresponds with a sharp tip. 

After extracting the coefficients from a fit, we can then simulate the following equation of motion using numerical continuation \cite{doedel2007auto} in order to obtain the nonlinear steady-state response of the driven system:

\begin{equation}
\begin{aligned}
\ddot{r} + \omega_0^2r + \frac{\omega_0}{Q}\dot{r} & - \frac{4\epsilon_{LJ}}{m} \left( -12 \frac{\sigma^{12}}{(r + d)^{13}} + 6 \frac{\sigma^6}{(r + d)^7} \right)  \\ 
& + \frac{C_1 V_{\rm DC}^2}{m(r + d)^2} = F_d\cos{\omega_d t}
\end{aligned}
\end{equation}

where $F_d$ is the drive level and $\omega_d$ is the drive frequency. We use a Q factor of $Q=10^6$, in line with the experiments, and set of drive levels $F_d$ according to the shifts of nonlinear resonance peaks with respect to the separation distance in the experiments. By setting different drive frequencies, we execute numerical continuation with separation distance, $d$, as the continuation parameter. This is done in order to simulate the Z-Sweep Resonance modality. \\
\begin{figure}[t]
    \centering
    \includegraphics[width=1\linewidth]{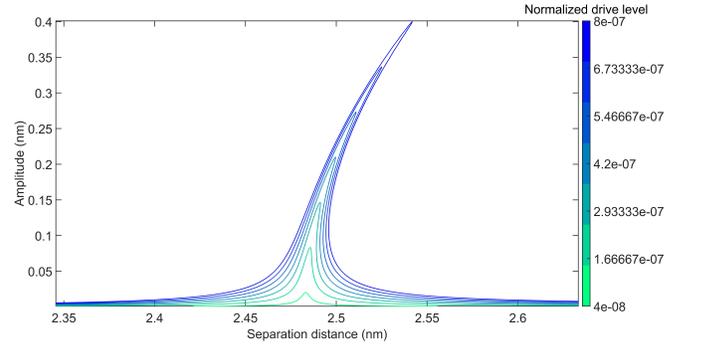}
    \caption{\textbf{Numerical continuation simulations.} Simulations of the system for $w_d/2\pi=277.77$ kHz, with different drive levels, showing the nonlinear steady-state response of the system for different separation distance.}
    \label{fig:NumCont}
\end{figure}

Numerical continuation simulations (Fig. \ref{fig:NumCont}) reveal multiple solutions for the steady-state of the system at high drive levels due to the nonlinear softening effect, similar to what happens in a Duffing oscillator. As the separation distance is increased during a Z-Sweep Resonance curve, it effectively increases the linear resonance frequency while the drive frequency is kept the same. This effectively imitates a frequency sweep in a reverse fashion (which is why we observe a shape similar to nonlinear hardening frequency response instead of a softening shape, in Fig. \ref{fig:NumCont}). The amplitudes found in the figure are added on top of the separation distance to help determine the effective distance for tunneling, allowing the curves in Fig.~\ref{fig:concept} to be established. As shown in Supplementary Note 9, a significant part of the current flows during the maximum extension of the oscillation.

\subsection{Membrane Navigation System}

\begin{figure*}[t]
    \includegraphics[width=\textwidth]{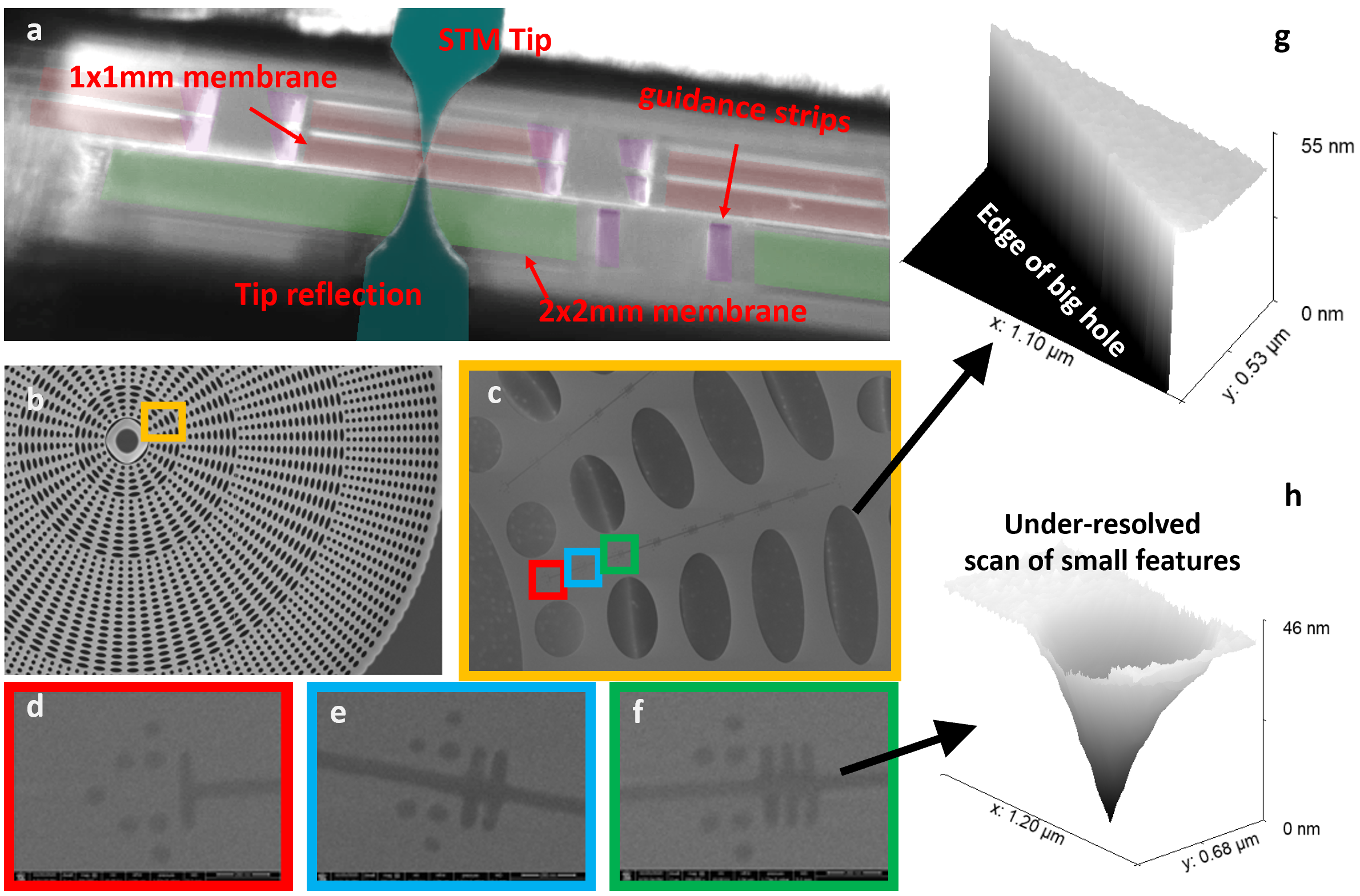}
    \caption{\textbf{Navigation System} \textbf{(a)} Optical image of the tip-sample junction, using false-color overlays for identification of various parts. \textbf{(b)} Electron Microscopy image of membrane with larger holes for underetching and small marker holes for navigation with a \textbf{(c)} zoom in and additional zoom ins \textbf{(d-f)}. \textbf{(g)} STM scan of large hole. Hole depth shows $55$~nm, but this was the maximum tip extension and no current was measured, indicating the true depth is larger. \textbf{(h)} STM scan of marker hole appearing as an poorly defined hole due to lack of tip optimization.}
    \label{fig:navigation}
\end{figure*}

The insets shown in Figure \ref{fig:xy-dependence}c-e are made possible thanks to optical access. Figure \ref{fig:navigation}a shows such image in more detail, highlighting an STM tip approaching from the top. A reflection of the tip aids in locating the probing point, with guiding strips on each membrane's side. Highlighted is also a $2 \times 2$~mm gold-coated membrane suspended over a gold-coated Si chip. The tip is shown probing a membrane coated by superconducting NbTiN on Si$_3$N$_4$. The two chips are placed firmly next to each other, allowing us to probe both types of membranes within a single experimental cycle. 

In earlier iterations of the experiment an attempt was made at measuring absolute out-of-plane motion of a membrane. To this end a membrane was made with a hole in the middle in which a pedestal was placed, such that the STM tip would measure the height difference between the pedestal and the oscillating membrane. In order to find this micrometer-sized pedestal, an on-membrane navigation system was implemented, see Figure \ref{fig:navigation}b-f. As with the membrane in this main work (see \textit{Sample Preparation} subsection), these earlier membranes had holes which helps with plasma etching. However, they served a double function as navigation system. The holes were laid out concentric around the pedestal, with an elliptical shape, the exact details of which depend on the relative distance and angle from the pedestal. \\

Additional markers with unique shapes were made to be smaller than $1 \times 1$~$\mu$m, as shown in panels d-f, such that a single marker would fit within a single scan frame of the STM. The experimentalist would then search their scan window for any such holes, determine its shape and deduce where to move from there. Due to the tip's apex shape the smaller markers were not properly resolved, but appeared as elliptic depressions (panel h). The large holes (used for underetching), on the other hand, resulted in a very large and very sudden height change (panel g). This allowed us to distinguish between both types of holes and help guide us towards the center of the membrane. \\

We did not optimize the tip apex for properly resolving the small markers. We expect STM tips that are prepared more carefully, such as through a combination of Field Ion Microscopy (FIM) and Field Emission Microscopy (FEM) to benefit more from this navigation system \cite{Gomer1961Harvard,Muller1969Elsevier}.

\subsection{Sample Preparation}
\textbf{Fabrication of the 2x2 mm$^2$ silicon nitride (Si$_3$N$_4$) mechanical resonator coated with gold:} a 340 nm Si$_3$N$_4$ film with 1.1 GPa tensile stress is deposited by low-pressure chemical vapor deposition (LPCVD) on a silicon wafer. The wafer is diced into 10 x 10 mm$^2$ chips. On a microchip, a square array of circular holes with radius 0.75 $\mu$m and center-to-center distance 3.5 $\mu$m in both x and y directions is patterned on the Si$_3$N$_4$ film. The circular holes are patterned by using electron-beam lithography on a positive tone ebeam resist (AR-P 6200), and subsequently transferred to the Si$_3$N$_4$ layer using inductively coupled plasma reactive ion etching  (ICP-RIE) based on CHF$_3$ and argon plasmas. The residual resist is removed by hot dimethylformamide (DMF), Then, the 2x2 mm$^2$ nanomechanical squared membranes is released by cryogenic SF$_6$ isotropic plasma etching the Si substrate using ICP etching at -120 $^{\circ}$C. Finally, a 45 nm gold film is coated on the suspended membrane, using thermal evaporation. In between the Si$_3$N$_4$ and gold films, a 2 nm thin chromium adhesive layer is added for coating.

\textbf{Fabrication of the 1x1 mm$^2$ Si$_3$N$_4$ mechanical resonator coated with niobium titanium nitride (NbTiN):} On a microchip coated with a 340 nm thick Si$_3$N$_4$ deposited the as described above, a 45 nm thick NbTiN layer is coated on it at 300 $^{\circ}$C using plasma-enhanced atomic layer deposition (PEALD). Then we pattern the circular holes the same ways as above, except that the array has a size of 1x1 mm$^2$, on the Si$_3$N$_4$/NbTiN bi-layer. At last, we suspend the 1x1 mm$^2$ squared membrane using cryogenic SF$_6$ isotropic plasma etching to undercut the Si substrate at -120 $^{\circ}$C.

\bibliography{bibliography}

@article{Halg2020PRA,
  title = {Membrane-Based Scanning Force Microscopy},
  author = {H\"alg, David and Gisler, Thomas and Tsaturyan, Yeghishe and Catalini, Letizia and Grob, Urs and Krass, Marc-Dominik and H\'eritier, Martin and Mattiat, Hinrich and Thamm, Ann-Katrin and Schirhagl, Romana and Langman, Eric C. and Schliesser, Albert and Degen, Christian L. and Eichler, Alexander},
  journal = {Phys. Rev. Appl.},
  volume = {15},
  issue = {2},
  pages = {L021001},
  numpages = {7},
  year = {2021},
  month = {Feb},
  publisher = {American Physical Society},
  doi = {10.1103/PhysRevApplied.15.L021001},
  url = {https://link.aps.org/doi/10.1103/PhysRevApplied.15.L021001}
}

@misc{Xu2025arXiv,
      title={Measuring Casimir Force Across a Superconducting Transition}, 
      author={Minxing Xu and Robbie J. G. Elbertse and Ata Keşkekler and Giuseppe Bimonte and Jinwon Lee and Sander Otte and Richard A. Norte},
      year={2025},
      eprint={2504.10579},
      archivePrefix={arXiv},
      primaryClass={quant-ph},
      url={https://arxiv.org/abs/2504.10579}, 
}

@book{Muller1969Elsevier,
  title={Field Ion Microscopy; Principles and Applications, by Erwin W. Muller and Tien Tzou Tsong},
  author={Muller, E.W. and Tsong, T.T.},
  lccn={10122495},
  url={https://books.google.com/books?id=htuS0AEACAAJ},
  year={1969},
  publisher={American Elsevier Publishing Company}
}

@book{Gomer1961Harvard,
  title={Field Emission and Field Ionization},
  author={Gomer, R.},
  isbn={9780608102641},
  lccn={60015237},
  series={Harvard monographs in applied science},
  url={https://books.google.com/books?id=zIh5AAAAIAAJ},
  year={1961},
  publisher={Harvard University Press}
}

@book{Israelachvili2011,
  author    = {Israelachvili, Jacob N.},
  title     = {Intermolecular and Surface Forces},
  edition   = {3},
  publisher = {Academic Press},
  address   = {Amsterdam},
  year      = {2011},
  isbn      = {978-0-12-391927-4},
  url={https://www.sciencedirect.com/book/monograph/9780123751829/intermolecular-and-surface-forces}
}

@Article{Thompson2008Nature,
author={Thompson, J. D.
and Zwickl, B. M.
and Jayich, A. M.
and Marquardt, Florian
and Girvin, S. M.
and Harris, J. G. E.},
title={Strong dispersive coupling of a high-finesse cavity to a micromechanical membrane},
journal={Nature},
year={2008},
month={Mar},
day={01},
volume={452},
number={7183},
pages={72-75},
issn={1476-4687},
doi={10.1038/nature06715},
url={https://doi.org/10.1038/nature06715}
}

@article{Reinhardt2016PRX,
  title = {Ultralow-Noise SiN Trampoline Resonators for Sensing and Optomechanics},
  author = {Reinhardt, Christoph and M\"uller, Tina and Bourassa, Alexandre and Sankey, Jack C.},
  journal = {Phys. Rev. X},
  volume = {6},
  issue = {2},
  pages = {021001},
  numpages = {8},
  year = {2016},
  month = {Apr},
  publisher = {American Physical Society},
  doi = {10.1103/PhysRevX.6.021001},
  url = {https://link.aps.org/doi/10.1103/PhysRevX.6.021001}
}

@article{Faust2012PRL,
  title = {Nonadiabatic Dynamics of Two Strongly Coupled Nanomechanical Resonator Modes},
  author = {Faust, Thomas and Rieger, Johannes and Seitner, Maximilian J. and Krenn, Peter and Kotthaus, J\"org P. and Weig, Eva M.},
  journal = {Phys. Rev. Lett.},
  volume = {109},
  issue = {3},
  pages = {037205},
  numpages = {4},
  year = {2012},
  month = {Jul},
  publisher = {American Physical Society},
  doi = {10.1103/PhysRevLett.109.037205},
  url = {https://link.aps.org/doi/10.1103/PhysRevLett.109.037205}
}

@article{Elfrink2009JMM,
doi = {10.1088/0960-1317/19/9/094005},
url = {https://doi.org/10.1088/0960-1317/19/9/094005},
year = {2009},
month = {aug},
publisher = {},
volume = {19},
number = {9},
pages = {094005},
author = {Elfrink, R and Kamel, T M and Goedbloed, M and Matova, S and Hohlfeld, D and van Andel, Y and van Schaijk, R},
title = {Vibration energy harvesting with aluminum nitride-based piezoelectric devices},
journal = {Journal of Micromechanics and Microengineering}
}

@article{Mahboob2008APL,
    author = {Mahboob, I. and Yamaguchi, H.},
    title = {Piezoelectrically pumped parametric amplification and Q enhancement in an electromechanical oscillator},
    journal = {Applied Physics Letters},
    volume = {92},
    number = {17},
    pages = {173109},
    year = {2008},
    month = {04},
    issn = {0003-6951},
    doi = {10.1063/1.2903709},
    url = {https://doi.org/10.1063/1.2903709},
}

@amisc{doedel2007auto,
	author = {Doedel, Eusebius J and Champneys, Alan R and Dercole, Fabio and Fairgrieve, Thomas F and Kuznetsov, Yuri A and Oldeman, B and Paffenroth, RC and Sandstede, B and Wang, XJ and Zhang, CH},
	publisher = {Concordia University},
	title = {AUTO-07P: Continuation and bifurcation software for ordinary differential equations},
	year = {2007},
    url={https://github.com/auto-07p/auto-07p}}

@article{norte2016,
  title={Mechanical resonators for quantum optomechanics experiments at room temperature},
  author={Norte, Richard A and Moura, Joao P and Gr{\"o}blacher, Simon},
  journal={Physical Review Letters},
  volume={116},
  number={14},
  pages={147202},
  year={2016},
  publisher={APS},
  url={https://doi.org/10.1103/PhysRevLett.116.147202}
}

@article{tsaturyan2017ultracoherent,
  title={Ultracoherent nanomechanical resonators via soft clamping and dissipation dilution},
  author={Tsaturyan, Yeghishe and Barg, Andreas and Polzik, Eugene S and Schliesser, Albert},
  journal={Nature Nanotechnology},
  volume={12},
  number={8},
  pages={776--783},
  year={2017},
  publisher={Nature Publishing Group UK London},
  url={https://doi.org/10.1038/nnano.2017.101}
}

@article{malcovati2018evolution,
  title={The evolution of integrated interfaces for MEMS microphones},
  author={Malcovati, Piero and Baschirotto, Andrea},
  journal={Micromachines},
  volume={9},
  number={7},
  pages={323},
  year={2018},
  publisher={MDPI},
  url={https://doi.org/10.3390/mi9070323}
}

@article{zawawi2020review,
  title={A review of MEMS capacitive microphones},
  author={Zawawi, Siti Aisyah and Hamzah, Azrul Azlan and Majlis, Burhanuddin Yeop and Mohd-Yasin, Faisal},
  journal={Micromachines},
  volume={11},
  number={5},
  pages={484},
  year={2020},
  publisher={MDPI},
  url={https://doi.org/10.3390/mi11050484}
}

@article{andrews2014bidirectional,
  title={Bidirectional and efficient conversion between microwave and optical light},
  author={Andrews, Reed W and Peterson, Robert W and Purdy, Tom P and Cicak, Katarina and Simmonds, Raymond W and Regal, Cindy A and Lehnert, Konrad W},
  journal={Nature Physics},
  volume={10},
  number={4},
  pages={321--326},
  year={2014},
  publisher={Nature Publishing Group UK London},
  url={https://doi.org/10.1038/nphys2911}
}

@article{thoen2016superconducting,
  title={Superconducting NbTin Thin Films With Highly Uniform Properties Over a 100 mm Wafer},
  author={Thoen, David Johannes and Bos, Boy Gustaaf Cornelis and Haalebos, EAF and Klapwijk, TM and Baselmans, JJA and Endo, Akira},
  journal={IEEE Transactions on Applied Superconductivity},
  volume={27},
  number={4},
  pages={1--5},
  year={2016},
  publisher={IEEE},
  url={https://doi.org/10.1109/TASC.2016.2631948}
}

@article{uder2018low,
  title={Low-force spectroscopy on graphene membranes by scanning tunneling microscopy},
  author={Uder, Bernd and Gao, Haibin and Kunnas, Peter and de Jonge, Niels and Hartmann, Uwe},
  journal={Nanoscale},
  volume={10},
  number={4},
  pages={2148--2153},
  year={2018},
  publisher={Royal Society of Chemistry},
  url={https://doi.org/10.1039/C7NR07300C}
}

@article{breitwieser2017,
  title={Investigating ultraflexible freestanding graphene by scanning tunneling microscopy and spectroscopy},
  author={Breitwieser, R and Hu, Yu-Cheng and Chao, Yen Cheng and Tzeng, Yi Ren and Liou, Sz-Chian and Lin, Keng Ching and Chen, Chih Wei and Pai, Woei Wu},
  journal={Physical Review B},
  volume={96},
  number={8},
  pages={085433},
  year={2017},
  publisher={APS},
  url={https://doi.org/10.1103/PhysRevB.96.085433}
}

@article{alyobi2020voltage,
  title={The voltage-dependent manipulation of few-layer graphene with a scanning tunneling microscopy tip},
  author={Alyobi, Mona M and Barnett, Chris J and Muratov, Cyrill B and Moroz, Vitaly and Cobley, Richard J},
  journal={Carbon},
  volume={163},
  pages={379--384},
  year={2020},
  publisher={Elsevier},
  url={https://doi.org/10.1016/j.carbon.2020.03.046}
}

@article{zan2012scanning,
  title={Scanning tunnelling microscopy of suspended graphene},
  author={Zan, Recep and Muryn, Chris and Bangert, Ursel and Mattocks, Philip and Wincott, Paul and Vaughan, David and Li, Xuesong and Colombo, Luigi and Ruoff, Rodney S and Hamilton, Bruce and others},
  journal={Nanoscale},
  volume={4},
  number={10},
  pages={3065--3068},
  year={2012},
  publisher={Royal Society of Chemistry},
  url={https://doi.org/10.1039/C2NR30162H}
}

@article{klimov2012electromechanical,
  title={Electromechanical properties of graphene drumheads},
  author={Klimov, Nikolai N and Jung, Suyong and Zhu, Shuze and Li, Teng and Wright, C Alan and Solares, Santiago D and Newell, David B and Zhitenev, Nikolai B and Stroscio, Joseph A},
  journal={Science},
  volume={336},
  number={6088},
  pages={1557--1561},
  year={2012},
  publisher={American Association for the Advancement of Science},
  url={https://doi.org/10.1126/science.1220335}
}

@article{xu2012atomic,
  title={Atomic control of strain in freestanding graphene},
  author={Xu, P and Yang, Yurong and Barber, SD and Ackerman, ML and Schoelz, JK and Qi, D and Kornev, Igor A and Dong, Lifeng and Bellaiche, L and Barraza-Lopez, Salvador and others},
  journal={Physical Review B—Condensed Matter and Materials Physics},
  volume={85},
  number={12},
  pages={121406},
  year={2012},
  publisher={APS},
  url={https://doi.org/10.1103/PhysRevB.85.121406}
}

@article{zhao2013fabrication,
  title={Fabrication and scanning tunneling microscopy characterization of suspended monolayer graphene on periodic Si nanopillars},
  author={Zhao, Xin and Zhai, Xiaofang and Zhao, Aidi and Wang, Bing and Hou, JG},
  journal={Applied physics letters},
  volume={102},
  number={20},
  year={2013},
  publisher={AIP Publishing},
  url={https://doi.org/10.1063/1.4807139}
}

@article{breitwieser2014flipping,
  title={Flipping nanoscale ripples of free-standing graphene using a scanning tunneling microscope tip},
  author={Breitwieser, Romain and Hu, Yu-Cheng and Chao, Yen Cheng and Li, Ren-Jie and Tzeng, Yi Ren and Li, Lain-Jong and Liou, Sz-Chian and Lin, Keng Ching and Chen, Chih Wei and Pai, Woei Wu},
  journal={Carbon},
  volume={77},
  pages={236--243},
  year={2014},
  publisher={Elsevier},
  url={https://doi.org/10.1016/j.carbon.2014.05.026}
}

@article{schoelz2015graphene,
  title={Graphene ripples as a realization of a two-dimensional Ising model: A scanning tunneling microscope study},
  author={Schoelz, JK and Xu, P and Meunier, V and Kumar, P and Neek-Amal, M and Thibado, PM and Peeters, FM},
  journal={Physical Review B},
  volume={91},
  number={4},
  pages={045413},
  year={2015},
  publisher={APS},
  url={https://doi.org/10.1103/PhysRevB.91.045413}
}

@article{uder2017convenient,
  title={A convenient method for large-scale STM mapping of freestanding atomically thin conductive membranes},
  author={Uder, B and Hartmann, U},
  journal={Review of Scientific Instruments},
  volume={88},
  number={6},
  year={2017},
  publisher={AIP Publishing},
  url={https://doi.org/10.1063/1.4985003}
}

@article{xu2024imaging,
  title={Imaging nanomechanical vibrations and manipulating parametric mode coupling via scanning microwave microscopy},
  author={Xu, Hao and Venkatachalam, Srisaran and Rabenimanana, Toky-Harrison and Boyaval, Christophe and Eliet, Sophie and Braud, Flavie and Collin, Eddy and Theron, Didier and Zhou, Xin},
  journal={Nano Letters},
  volume={24},
  number={28},
  pages={8550--8557},
  year={2024},
  publisher={ACS Publications},
  url={https://doi.org/10.1021/acs.nanolett.4c01401}
}

@article{Hwang2022RevSciIns,
    author = {Hwang, Jiyoon and Krylov, Denis and Elbertse, Robbie and Yoon, Sangwon and Ahn, Taehong and Oh, Jeongmin and Fang, Lei and Jang, Won-jun and Cho, Franklin H. and Heinrich, Andreas J. and Bae, Yujeong},
    title = "{Development of a scanning tunneling microscope for variable temperature electron spin resonance}",
    journal = {Review of Scientific Instruments},
    volume = {93},
    number = {9},
    pages = {093703},
    year = {2022},
    month = {09},
    issn = {0034-6748},
    doi = {10.1063/5.0096081},
    url = {https://doi.org/10.1063/5.0096081},
}

@article{Bettac2009Nanotech,
doi = {10.1088/0957-4484/20/26/264009},
url = {https://dx.doi.org/10.1088/0957-4484/20/26/264009},
year = {2009},
month = {jun},
publisher = {},
volume = {20},
number = {26},
pages = {264009},
author = {Andreas Bettac and Juergen Koeble and Konrad Winkler and Bernd Uder and Markus Maier and Albrecht Feltz},
title = {QPlus: atomic force microscopy on single-crystal insulators with small oscillation
amplitudes at 5 K},
journal = {Nanotechnology}
}

\clearpage

\onecolumngrid

\setcounter{section}{0}
\setcounter{figure}{0}
\setcounter{table}{0}
\setcounter{equation}{0}

\renewcommand{\thesection}{S\arabic{section}}
\renewcommand{\thefigure}{S\arabic{figure}}
\renewcommand{\thetable}{S\arabic{table}}
\renewcommand{\theequation}{S\arabic{equation}}


\title{Detection of MEMS Acoustics via Scanning Tunneling Microscopy (Supplementary)}

\author{R.\ J.\ G.\ Elbertse\textsuperscript{\S}}
\affiliation{Kavli Institute of Nanoscience, Department of Quantum Nanoscience, Delft University of Technology, Delft, The Netherlands}

\author{M.\ Xu\textsuperscript{\S}}
\affiliation{Kavli Institute of Nanoscience, Department of Quantum Nanoscience, Delft University of Technology, Delft, The Netherlands}
\affiliation{Department of Precision and Microsystems Engineering, Delft University of Technology, Delft, The Netherlands}

\author{A.\ Ke\c{s}kekler}
\affiliation{Department of Precision and Microsystems Engineering, Delft University of Technology, Delft, The Netherlands}

\author{S.\ Otte}
\affiliation{Kavli Institute of Nanoscience, Department of Quantum Nanoscience, Delft University of Technology, Delft, The Netherlands}

\author{R.\ A.\ Norte}
\affiliation{Department of Precision and Microsystems Engineering, Delft University of Technology, Delft, The Netherlands}
\affiliation{Kavli Institute of Nanoscience, Department of Quantum Nanoscience, Delft University of Technology, Delft, The Netherlands}

\maketitle

\vspace{-6pt}
\begin{center}
\footnotesize \textsuperscript{\S}These authors contributed equally to this work.
\end{center}
\vspace{12pt}

\tableofcontents
\clearpage

\section{Part 1: Figure Settings}
Table \ref{tab:params} describes all the major settings relevant for the figures in the main text and supplementary information. Besides describing the used modality, this table includes the tunneling current setpoint used for the feedback loop $I$. In case of Z-Sweep Resonance this means the current at $z = 0$~nm. Shown is also the bias voltage $V_{\rm DC}$ at which the current setpoint is defined. For the LCPD measurements these settings are used to define a common starting point for each curve, prior to adjusting the voltage. The table also shows the temperature at which the data was taken and the coating of the membrane, as well as the mode being measured. The RF generator's output settings such as the power $P$ and the frequency $f$ are given as well.

\definecolor{z-sweep}{HTML}{D44CD1}
\definecolor{feedback}{HTML}{BA8600}
\definecolor{homodyne}{HTML}{4472C4}

\begin{table}[h]
\centering
\footnotesize
\caption{Summary of experimental parameters. HD stands for Homodyne Detection, FR stands for Feedback Resonance and Z-SR stands for Z-Sweep Resonance.}

\setlength{\tabcolsep}{5pt}
\renewcommand{\arraystretch}{1.2}

\begin{tabular}{|l|c|c|c|c|c|c|c|c|}
\hline
Figure
& Modality
& $I$ (pA)
& $V_{\mathrm{DC}}$ (mV)
& $T$ (K)
& Membrane
& Mode
& $P$ (dBm)
& $f$ (kHz) \\
\hline
Fig.~1c & \textcolor{homodyne}{HD} & 25 & 30 & 4.6 & Au & (2,2) & -30 & [299.7, 299.8] \\
Fig.~1d & \textcolor{feedback}{FR} & 12.5 & 50 & 4.6 & NbTiN & (1,2) & -40 & [435.9,  436.1] \\
Fig.~1e & \textcolor{z-sweep}{Z-SR} & 40 & 50 & 4.6 & NbTiN & (1,1) & -90 & 276.956 \\
Fig.~2c & \textcolor{z-sweep}{Z-SR} & 100 & 50 & 4.6 & NbTiN & (1,1) &-80/-85/-90 & 278.610 / 278.615\\
Fig.~3a & \textcolor{z-sweep}{Z-SR} & 10 & 50 & 4.6 & NbTiN & (1,1) & -75 & 277.75  \\
3a inset & \textcolor{z-sweep}{Z-SR} & 10 & 50 & 4.6 & NbTiN & (1,1) & -75 & [277.748, 277.763] \\
Fig.~3b & \textcolor{feedback}{FR} & 10 & 50 & 4.6 & NbTiN & (1,1) & -90 & [277.73, 277.78] \\
Fig.~4a & \textcolor{feedback}{FR} & 12.5 & 100 & 10.2 & NbTiN & (1,2) & -55 & [435.8,  436.1] \\
Fig.~4c-e & \textcolor{homodyne}{HD} & 12.5 & 50 & 10.2 & Au & various & -30 & [100, 1000] \\
Fig.~5a & \textcolor{z-sweep}{Z-SR} & 10 & 10/1000 & 4.6 & NbTiN & (1,1) & various & [277, 277.79865] \\
Fig.~5b & \textcolor{z-sweep}{Z-SR} & 10 & 10/1000 & 4.6 & NbTiN & (1,1) & -50 & 277.7978\_ \\
\hline
Fig.~S1a & feedback & 2 & 50 & 4.6 & NbTiN & (1,1) & -50 & 277.7748 \\
Fig.~S1b & feedback & 2/3/5/10/50/100 & 50 & 4.6 & NbTiN & (1,1) & -50 & various \\
Fig.~S1b & feedback & 1/2/50 & 50 & 4.6 & NbTiN & (2,1) & -30 & various \\
Fig.~S2 & spectrum & 1400 & 10 & 4.6/7.0/10.2 & NbTiN & \textbf{---} & \textbf{---} & \textbf{---} \\
Fig.~S3b-e & \textcolor{homodyne}{HD} & 12.5 & 50 & 4.6 & Au & (2,2) & -30 & [295, 305] \\
Fig.~S3f & \textcolor{homodyne}{HD} & 12.5 & 50 & 4.6 & NbTiN & (1,1)/(1,2)/(2,1) & -20 & [300, 500] \\
Fig.~S4a-c & \textcolor{homodyne}{HD} / \textcolor{feedback}{FR} & 12.5 & 50 & 4.6 & NbTiN & (1,2) & -50/-45/-40 & [435.5, 436.5] \\
Fig.~S4d-f & \textcolor{homodyne}{HD} / \textcolor{feedback}{FR} & 12.5/100/500 & 50 & 4.6 & NbTiN & (1,2) & -47.5 & [435.5, 436.5] \\
Fig.~S5a & \textcolor{feedback}{FR} & 100 & 50 & 4.6 & NbTiN & (1,2) & -47.5 & [435.8, 436.0] \\
Fig.~S5b & \textcolor{feedback}{FR} & 12.5 & 50 & 4.6 & NbTiN & (1,2) & -50 & [436.0, 435.8] \\
Fig.~S5c & \textcolor{feedback}{FR} & 675 & 50 & 4.6 & NbTiN & (1,1) & -75 & [278.490, 278.515] \\
Fig.~S5d & \textcolor{feedback}{FR} & 100 & 50 & 4.6 & NbTiN & (1,1) & -110/-112/-114 & [275.870, 275.875] \\
Fig.~S6a & feedback & various & -100 & 4.6 & NbTiN & \textbf{---} & \textbf{---} & \textbf{---} \\
Fig.~S6b & feedback & various & -150 & 4.6 & NbTiN & \textbf{---} & \textbf{---} & \textbf{---} \\
Fig.~S6c & feedback & various & various & 4.6 & NbTiN & \textbf{---} & \textbf{---} & \textbf{---} \\
Fig.~S7a & \textcolor{feedback}{FR} & 5, 10 & various & 4.6 & NbTiN & (1,1) & -70 & [279.386, 279.546] \\
Fig.~S7b & \textcolor{feedback}{FR} & 5 & various & 4.6 & NbTiN & (1,1) & -70, -75 & [279.5256, 279.546] \\
Fig.~S7c & \textcolor{feedback}{FR} & 5 & various & 4.6 & NbTiN & (1,1) & -75, -80 & [277.738, 277.748] \\
Fig.~S8a/b & \textcolor{z-sweep}{Z-SR} & 10 & 10 & 4.6 & NbTiN & (1,1) & -50 & 277.415 \\
Fig.~S8c/d & \textcolor{z-sweep}{Z-SR} & 10 & 1000 & 4.6 & NbTiN & (1,1) & -60 & 277.77 \\
Fig.~S8e/f & \textcolor{z-sweep}{Z-SR} & 10 & 1000 & 4.6 & NbTiN & (1,1) & -50 & 277.79784 \\
Fig.~S8 & \textcolor{z-sweep}{Z-SR} & 10 & 10/1000 & 4.6 & NbTiN & (1,1) & various & [277, 277.79865] \\
\hline
\end{tabular}\label{tab:params}

\end{table}

\section{Part 2: Q-factor}
\begin{figure}[ht]
    \centering
    \includegraphics[width=\textwidth]{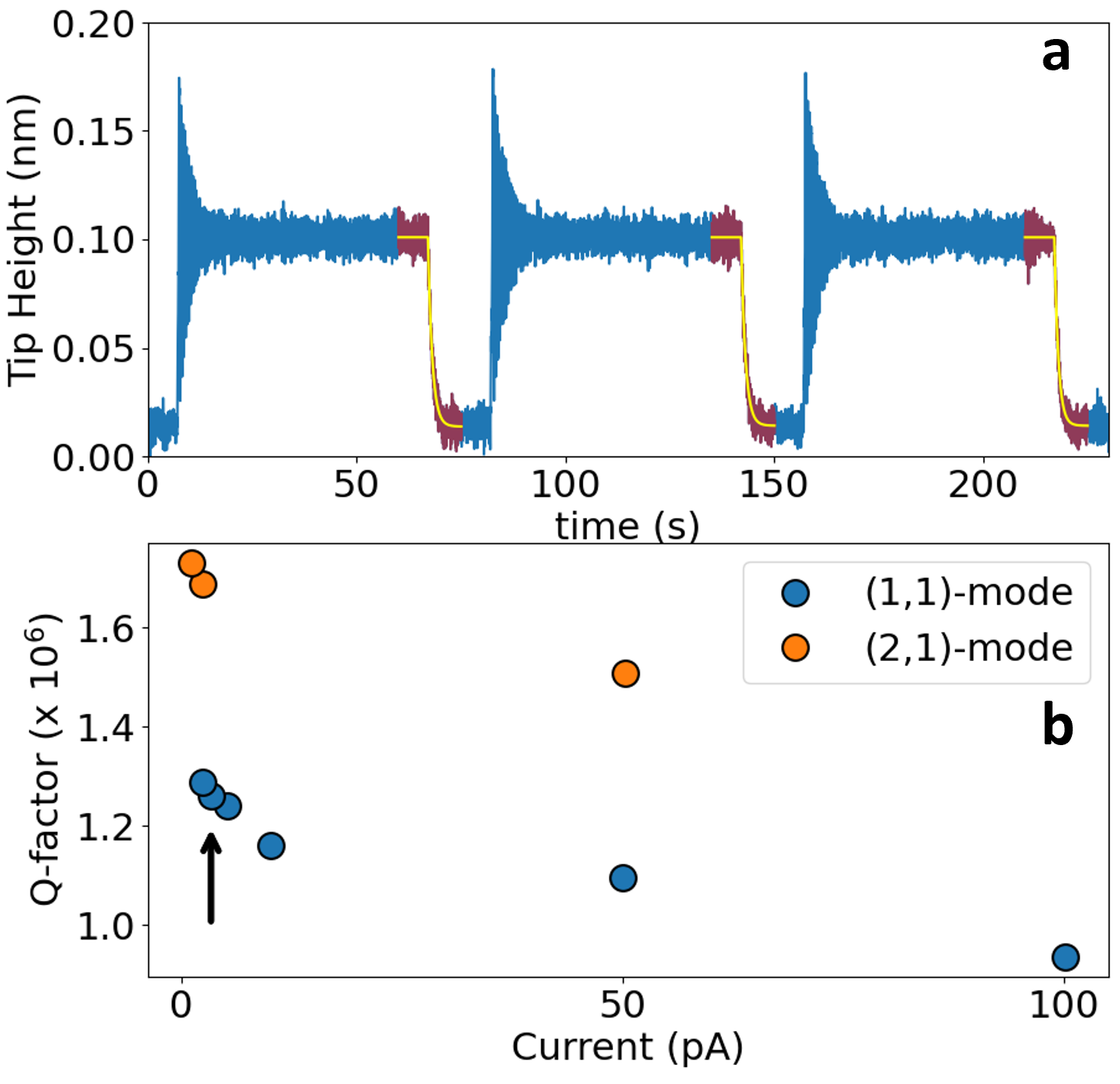}
    \caption{\textbf{Q-factor} \textbf{(a)} Tip height traces, applying a driving power $P = -50$~dBm after a few second at a current setpoint of $10$~pA. Around $60$~s the driving power is cancelled and the membrane relaxes, resulting in a decaying amplitude whose envelope is matched by the tip height. Yellow lines indicate fit to exponential decay. \textbf{(b)} Q-factor for various current setpoints and for two different modes. Arrow indicates data shown in panel \textbf{a}.}
    \label{fig:Q}
\end{figure}

To determine the Q-factor of the membrane and the influence of the tip, we performed ringdown measurements. These are performed by applying a driving voltage while feedback is on long enough until the system is in equilibrium and then turning off the driving voltage. As the amplitude decays over time, the tip will adjust its height accordingly to maintain a stable current. With the feedback response being much faster (order miliseconds) than the ringdown time (order seconds), this is a valid method of probing the Q-factor. This is shown in Fig.~S\ref{fig:Q}a where, after maintain the driving voltage for about 60 seconds, a drop in the tip height is observed upon turning off the driving voltage. A range before and after this drop, shown in red, is used to fit the drop to an exponential decay with decay time $t_0$. This process is repeated about $10$ times and an average $t_0$ is determined (two repetitions are shown). From this average decay time the Q-factor is derived.\\

The Q-factor is determined following this procedure for various current setpoints, and for both the (1,1)-mode and the (2,1)-mode of the NbTiN-coated membrane, see panel b. As expected, when the current setpoint is higher, i.e. when the tip is closer to the membrane, the Q-factor is lower indicating a tip-induced damping. We also found the higher mode to have a larger Q-factor. 

\section{Part 3: Scanning Tunneling Spectroscopy of Superconducting Gap}

\begin{figure}[ht]
    \centering
    \includegraphics[width=\textwidth]{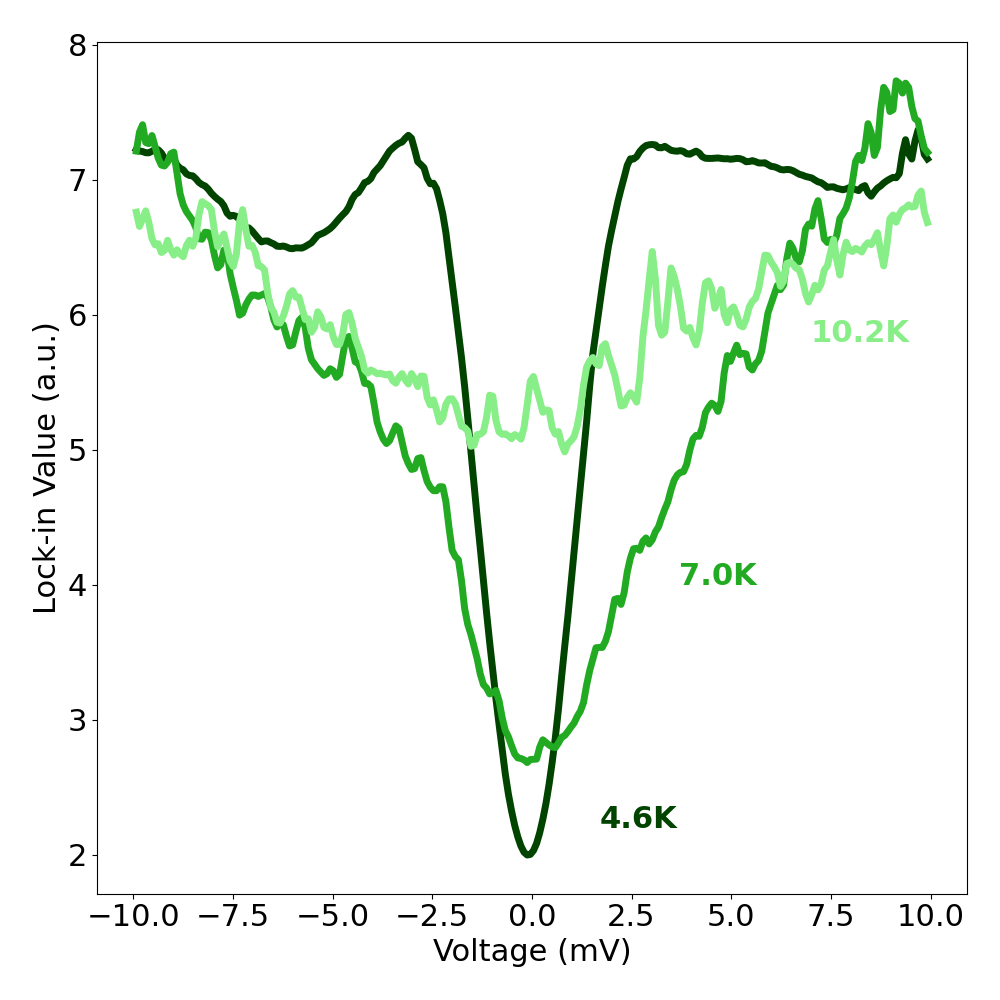}
    \caption{\textbf{Scanning Tunneling Spectroscopy.} Measured density of states of superconducting gap at three different temperatures. The higher temperatures broaden the superconducting gap.}
    \label{fig:SC}
\end{figure}

To demonstrate that our setup can probe the electronic states of our membrane, we performed bias spectroscopy measurements  across the junction between the tip and the NbTiN-coated membrane (Fig. S\ref{fig:SC}). The spectra exhibit the expected broadening of the superconducting gap and reduction in peak sharpness as the temperature is increased from 4.6 K to 10.2 K, consistent with standard BCS behavior. 

\section{Part 4: Signal chopping}

\begin{figure}[ht]
    \centering
    \includegraphics[width=\textwidth]{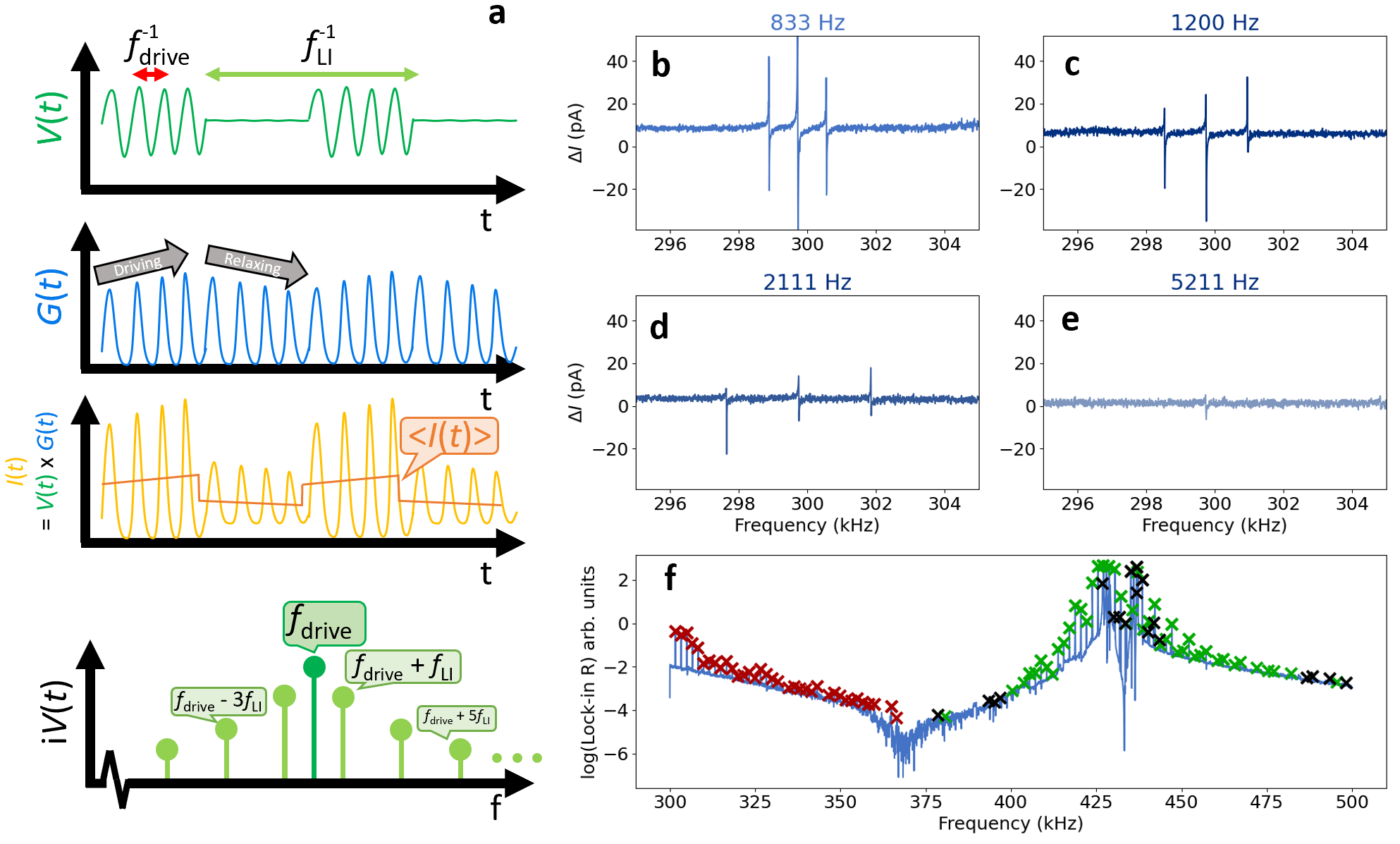}
    \caption{\textbf{Homodyne Detection} \textbf{(a)} [Top graph] Applied voltage to the tip-membrane junction, consisting of a DC voltage with a driving voltage oscillating at $f_{\rm drive}$. The driving voltage is amplitude modulated (chopped with a $50\%$ duty cycle) at the lock-in frequency $f_{\rm LI}$. When $f_{\rm drive}$ is close to the resonance frequency of the membrane, the resulting conductance [Second graph] of the tip-membrane junction follows a similar oscillatory behavior, but enhanced due to the exponential current dependence of the tunnel junction. When the driving is "on" the membrane's amplitude increases, during the off-cycle it will relax. [Third graph] With the current as the product of voltage and conductance, this, too, will follow an oscillatory behavior. The time-averaged conductance $\langle I(t) \rangle$, which will be the only measurable current, will be higher during the on-cycle than during the off-cycle. [Bottom graph] Taking a Fourier Transform of $V(t)$ results in a peak at $f_{\rm drive}$ with side-peaks spaced $2f_{\rm LI}$ apart. \textbf{(b-e)} Lock-in measurements showing two prominent side-peaks, which move further out as $f_{\rm LI}$ increases. The peak height also decreases significantly as $f_{\rm LI}$ increases, owing to the low pass filter behavior of the pre-amplifier. \textbf{(f)} Frequency sweep from $300$~kHz to $500$~kHz with a large driving power of $-20$~dBm showing the logarithm of the lock-in R. Sidepeaks of three eigenmodes are observed, one at around $280$~kHZ ((1,1)-mode, red crosses), and two at $428$~kHz and $436$~kHz ((2,1)- and (1,2)-modes, green and black crosses). Measurements are taken on NbTiN-coated membrane.}
    \label{fig:homodyne}
\end{figure}

The Homodyne Detection modality relies on a fixed phase relation between the applied oscillating voltage and resulting conductance. This leads both to an enhanced oscillating current, as well as a phase-dependent DC current offset, known as rectification. The enhanced oscillating current is not directly measurable, as the preamplifier limits the temporal resolution to $1.1$~kHz, whereas the driving frequency $f_{\rm drive}$ operates at several hundred kHz. The rectified signal is measurable, albeit often below the noise limit. To overcome these limitations,  the signal is amplitude-modulated (chopped with $50\%$ duty cycle) at a frequency of $f_{\rm LI} = 833$~Hz, as shown in Fig.~S\ref{fig:homodyne}a (top). When the oscillation voltage is “on,” the membrane is driven resonantly and its mechanical oscillation amplitude increases, whereas when the oscillating voltage is “off,” the mechanical amplitude decays. Thanks to the exceptionally high Q-factor of about $10^6$, the membrane’s amplitude decays only slowly compared to the chopping frequency, allowing it to reach a steady-state equilibrium between resonant driving and relaxation. The tip-sample conductance is dependent on the tunneling distance, i.e. the tip-membrane distance, with the conductance generally following $G = G_0 e^{-z/z_c}$, where $G_0$ is a conductance coefficient, $z$ is the tunneling distance and $z_c$ is the tunneling constant. The second graph in panel a shows the resulting conductance, which peaks when the tip and membrane are nearby and dips when they are far apart. The maximum conductance depends on the mechanical oscillation amplitude. The third graph shows that the majority of current flows when the tip-membrane distance is smallest, as also derived in Supplementary Note 9. During the on-cycles these moments of high conductance coincide with a higher voltage, resulting in an overall enhanced current. During the off-cycle, no such enhancement takes place, resulting in an overall lower current. These two different current values alternate at the chopping frequency $f_{\rm LI}$, and is sent to a lock-in amplifier operating at the same frequency. The lock-in amplified signal is less sensitive to noise than any other signal, and thus is able to detect even very small variations in current as a result of an oscillating conductance in-phase with an applied oscillating voltage. \\

Since the lock-in signal is proportional to the driving voltage, it becomes impossible to register a lock-in signal for very low driving powers, even if the membrane does respond very strongly. For this reason we have not found any lock-in signal for driving voltages below $-66$~dBm. We also found that only this modality worked on the gold-coated membrane, which we interpret as being the result of its lower compliance, as this modality can still measure sub-pm mechanical amplitudes.  \\

Following Fourier analysis of a chopped sine wave, see the bottom graph of Fig.~\ref{fig:homodyne}a, the input voltage is actually composed of an infinite amount of sine waves:

\begin{equation}\label{eq:Fourier}
 V = V_{\rm DC} + \frac{V_0}{2} \sin(f_{\rm drive} t)) + \sum_{i=-\infty}^{\infty} \frac{V_0}{i} \sin(((1+2i)f_{\rm LI} + f_{\rm drive})  t)   
\end{equation}

Thus, one would expect to see resonance responses not only at $f_{\rm drive}$ but at each of the other associated frequencies as well. Figure S\ref{fig:homodyne}b shows the associated homodyne peaks for $f_{\rm LI} = 833$~Hz, $1200$~Hz, $2111$~Hz and $5211$~Hz as measured on a NbTiN-coated membrane. It also shows that, as the lock-in frequency increases beyond $1100$~Hz, the signal strength is significantly reduced, consistent with the preamplifier's low-pass filter behavior. \\

One could even make use of these side-peaks to help find the resonance frequency. Figure S\ref{fig:homodyne}c shows a frequency sweep between $300$~kHz and $500$~kHz, containing side-peaks of three main resonances, one around $300$~kHz ((1,1)-mode), one around $428$~kHz ((2,1)-mode) and one around $436$~kHz ((1,2)-mode). These three families of side-peaks are marked with red, green and black crosses, respectively. Even more than $50$~kHz away from their main resonance is it still possible to find side-peaks. \\

\section{Part 5: Linear Regime}
\begin{figure}
    \centering
    \includegraphics[width=\textwidth]{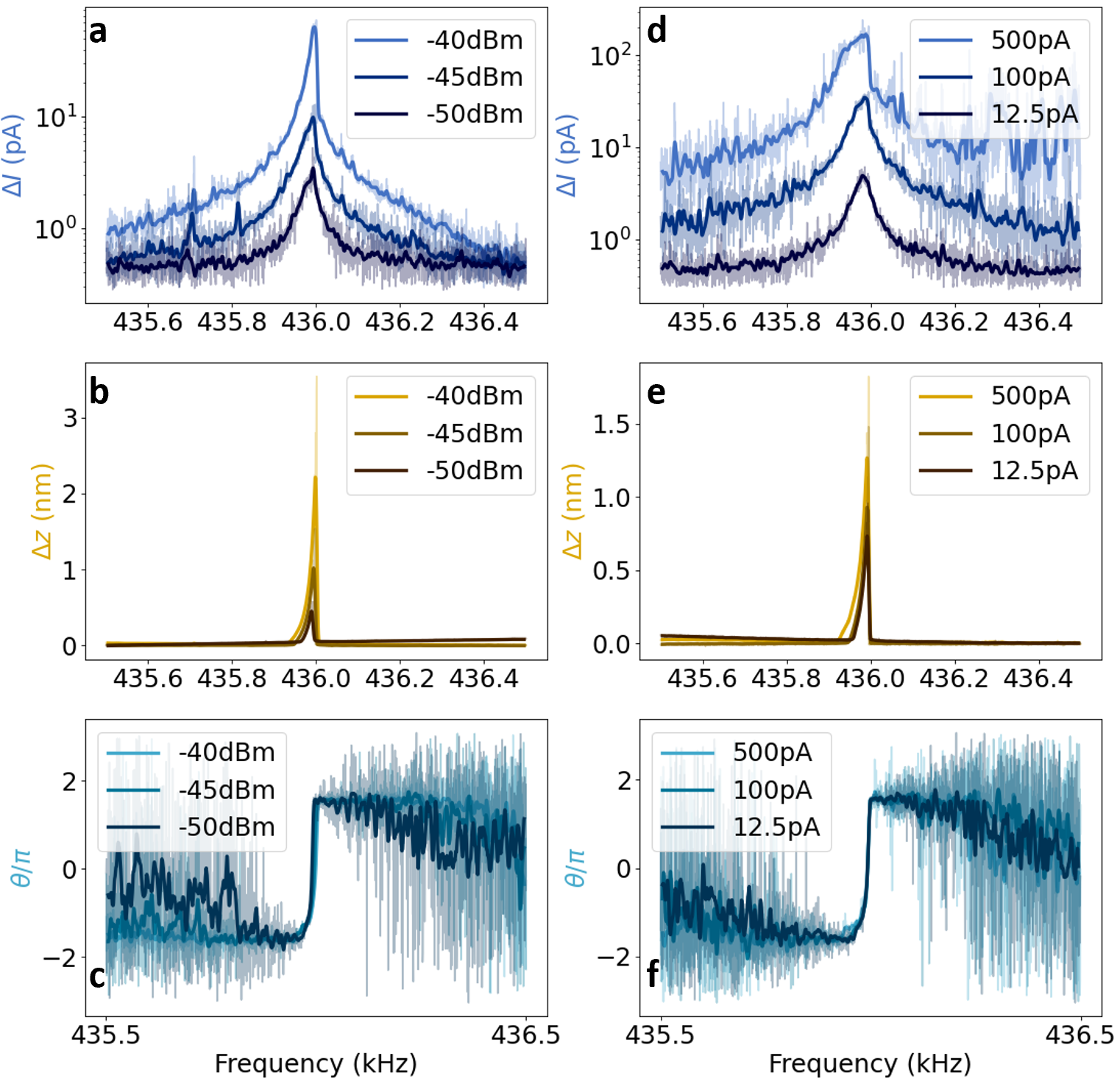}
    \caption{ \textbf{Various measurement channels} \textbf{(a)} Homodyne detection response for three different power settings. \textbf{(b)} Same data as panel b, but showing the feedback resonance response. \textbf{(c)} Same data as panel c but showing the phase of the homodyne signal. \textbf{(d-f)} Same as panels a-c, but for three different data sets, taken at different current setpoints.}
    \label{fig:linear}
\end{figure}

Figure 1e of the main text shows a sudden drop in the current during the Z-Sweep Resonances, which occurrs from the non-linear response of the membrane to the applied forces. The drop in tip height for the Feedback Resonances, shown in Figure 1d of the main text, is the result of a similar response mixed with a feedback cycle. Here we will explore this non-linear response further, including also Homodyne Detection simultaneously with Feedback Resonance and show that the drops occuring with Feedback Resonance modality even occur when the membrane is operating in the linear regime. Figure S\ref{fig:linear} explores this for different driving powers (panels a-c) and different tunnel currents at constant voltage (panels d-f). The first row of panels shows the homodyne amplitude signal, the second row shows the tip height, and the third row shows the lock-in phase of the signal of the first row. \\

Unsurprisingly we find that a larger current or applied oscillating voltage leads to larger homodyne signal (panels a and d). Not only is the amplitude larger, but the signal to noise ratio is larger than unity for a bigger range, when either increases. This results also in the phase of the signal to be better defined further away from the peak (panels c and f). Increasing the oscillating voltage amplitude also leads to a larger frequency window over which the mechanical oscillations are large enough to induce a tip retraction (panel b), consistent with panel a. This is less so the case when the current increases, as the feedback settings take into account the higher current. Nonetheless, panel e shows that the $500$~pA sweep yields a tip retraction over a larger frequency range, suggesting a larger oscillating electric field between tip and membrane. \\

We note that the homodyne signal indicates a linear response for the darkest data sets shows in panels a and d. In that case, the sudden drop shown in panels b and e, for the same settings, may be interpreted as due a more complex interplay between $f_{\rm drive}$, the shifting $f_0$ and how the constant current feedback loop interacts with that.

\section{Part 6: Feedback Oscillations and Telegraphic Noise} 

\begin{figure}[ht]
    \centering
    \includegraphics[width=\textwidth]{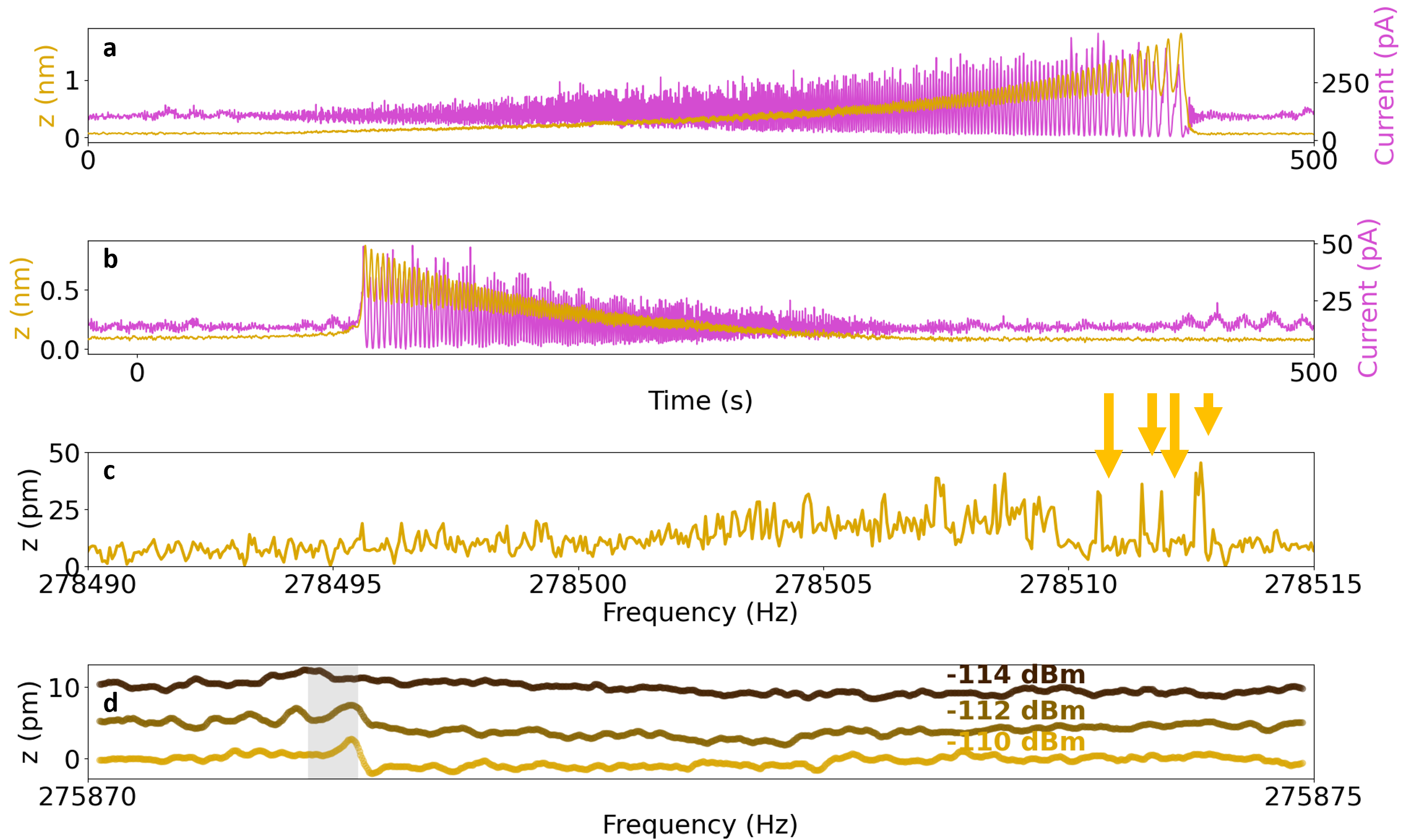}
    \caption{\textbf{Feedback Oscillations and Telegraphic Noise} \textbf{(a)} Measured current and tip height $z$ during a feedback resonance measurement. Oscillations occur as the feedback cycle is not able to properly keep up to the changing measured current. These oscillations can be avoided by picking better feedback parameters. \textbf{(b)} Same as a but for a backwards sweep. \textbf{(c)} Feedback resonance measurement showing telegraphic noise, indicated by the arrows. \textbf{(d)} Feedback resonance measurement with driving powers of $P = -110$~dBm, $-112$~dBm and $-114$~dBm, showing a sudden drop-off which corresponds to $f_{\rm drop}$. Note that due to the amplitude modulation (see Supplementary Note 4) the actual driving power is $3$~dBm less.}
    \label{fig:oscillation}
\end{figure}

During the Feedback Resonance modality, ideally the feedback loop is able to maintain the proper tip-membrane distance to yield a stable current, as read by the low-pass filter of the preamplifier at $1.1$~kHz. If the feedback settings are not set properly, the tip height turns unstable, resulting in large oscillations, as evidenced during both a forward sweep (Fig.~\ref{fig:oscillation}a) and a backward sweep (Fig.~\ref{fig:oscillation}b). Since the Q-factor is large enough ($Q > 10^6$, see Supplementary Note 2), these oscillations in tip position, with period around $100$~ms, are fast enough to not disturb the overall Feedback Resonance measurement. While sufficient integration time could obscure these oscillations in the data, all other presented data does not have such oscillations as we further optimized the feedback settings.\\

During some Feedback Resonance measurements, especially as the tip is close, we observed telegraphic noise, as shown in Fig.~\ref{fig:oscillation}c, indicating the membrane is switching between two meta-stable states. This is consistent with our simulations, shown in the Methods section, where sufficient driving forces causes the resonance peak to bend. \\

Finally, we show the lowest driving power with which we managed to observe a resonance. In this case, the observation follows from the sudden drop in tip height during a Feedback Resonance measurement, as shown in Fig.~\ref{fig:oscillation}d. Since we always amplitude modulate (i.e. chop our signal) to acquire a Homodyne signal where possible, the real driving power is $3$~dB less. Therefore, we find evidence of resonance even at an effective driving power of $-117$~dBm, or $2$~fW.

\section{Part 7: Stability}
\begin{figure}[ht]
    \centering
    \includegraphics[width=\textwidth]{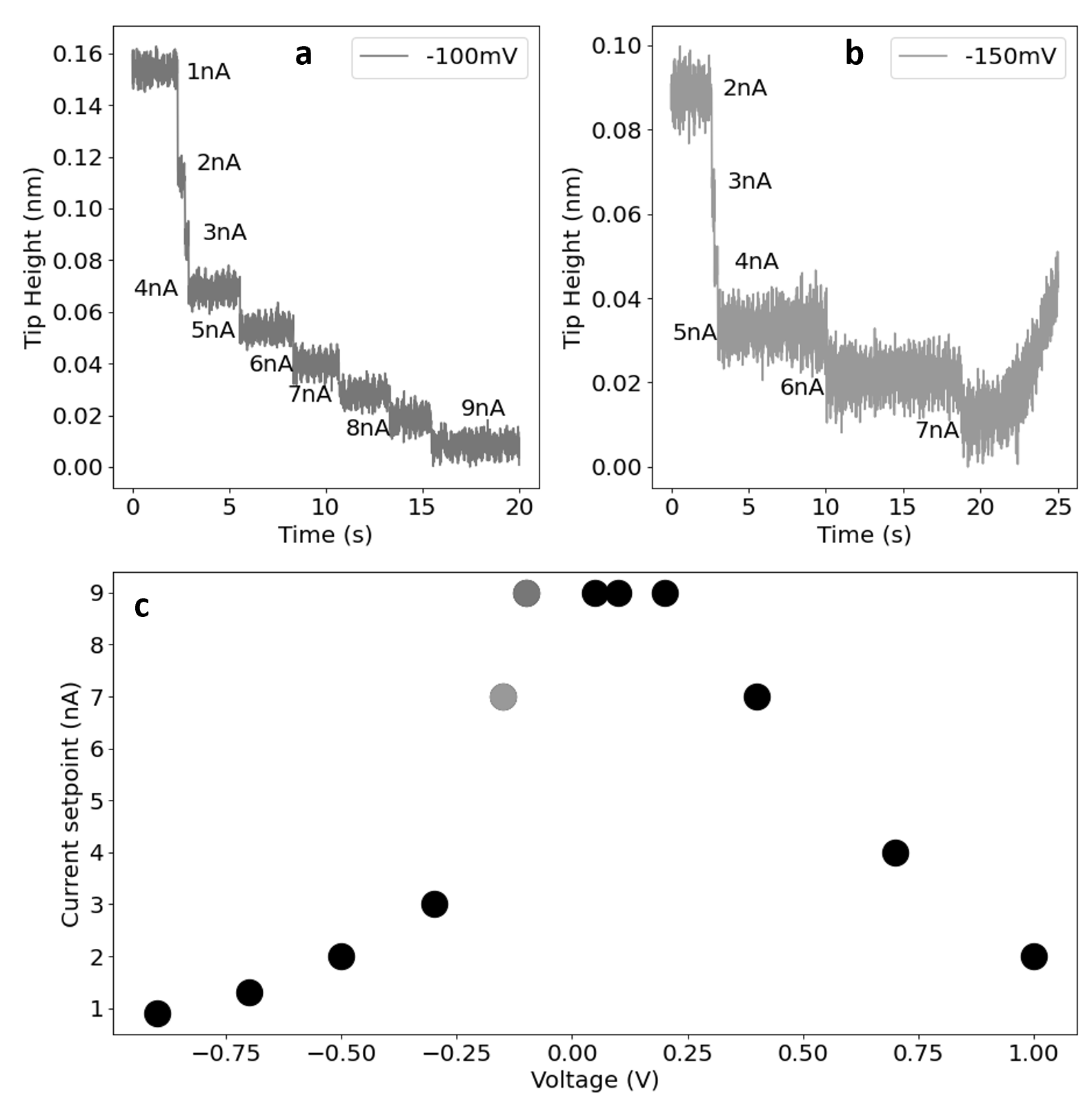}
    \caption{\textbf{Stability }\textbf{(a)} Tip height with feedback on, at $-100$~mV. Over time the current setpoint is manually increased, resulting in a lower tip height. \textbf{(b)} Same as \textbf{a} but taken at $-150$~mV and an unstable behavior observed when the current setpoint is set to $7$~nA. \textbf{(c)} Overview of unstable setpoints. Values at $9$~nA indicate stable positions observed up to this value, whereas values below this indicate unstable values observed. }
    \label{fig:stability}
\end{figure}

We found that under certain conditions, the feedback loop to maintain a constant current becomes unstable. We found that this is dependent on both the applied voltage and the current setpoint. Figure S\ref{fig:stability}a shows that when a bias voltage of $-100$~mV is applied between the tip and membrane, the feedback controller is always able to maintain a steady position. We manually increase the setpoint and found this to be true for currents up to at least $9$~nA. We did not explore higher currents as the preamplifier at a gain of $10^9$ is limited to signals up to $10$~nA. Figure S\ref{fig:stability}b, however, shows that when $V_{\rm DC} = -150$~mV is applied, the tip is automatically retracting when a setpoint of $7$~nA is applied. While the retraction is slow, it nonetheless signifies instability. We did not apply any driving power during these tests.  \\

Exploring this stability for a larger range of bias voltages, Fig.~S\ref{fig:stability} shows the highest current setpoint we managed to maintain stable conditions. Values of $9$~nA indicate the tip was stable even at that value, whereas the other data points mark the lowest current in which instability was found. The two data points of panels a and b reflect this, with the lighter colored circles. Noting that at a higher voltage one is able to maintain the same current at a larger distance, it appears that the onset of instability is \textit{not} dependent on only the tip-membrane distance alone but really depends on the applied voltages as well. The data was taken with the same tip apex as Figure 3 in the main text.

\section{Part 8: LCPD Variations}
\begin{figure}[ht]
    \centering
    \includegraphics[width=\textwidth]{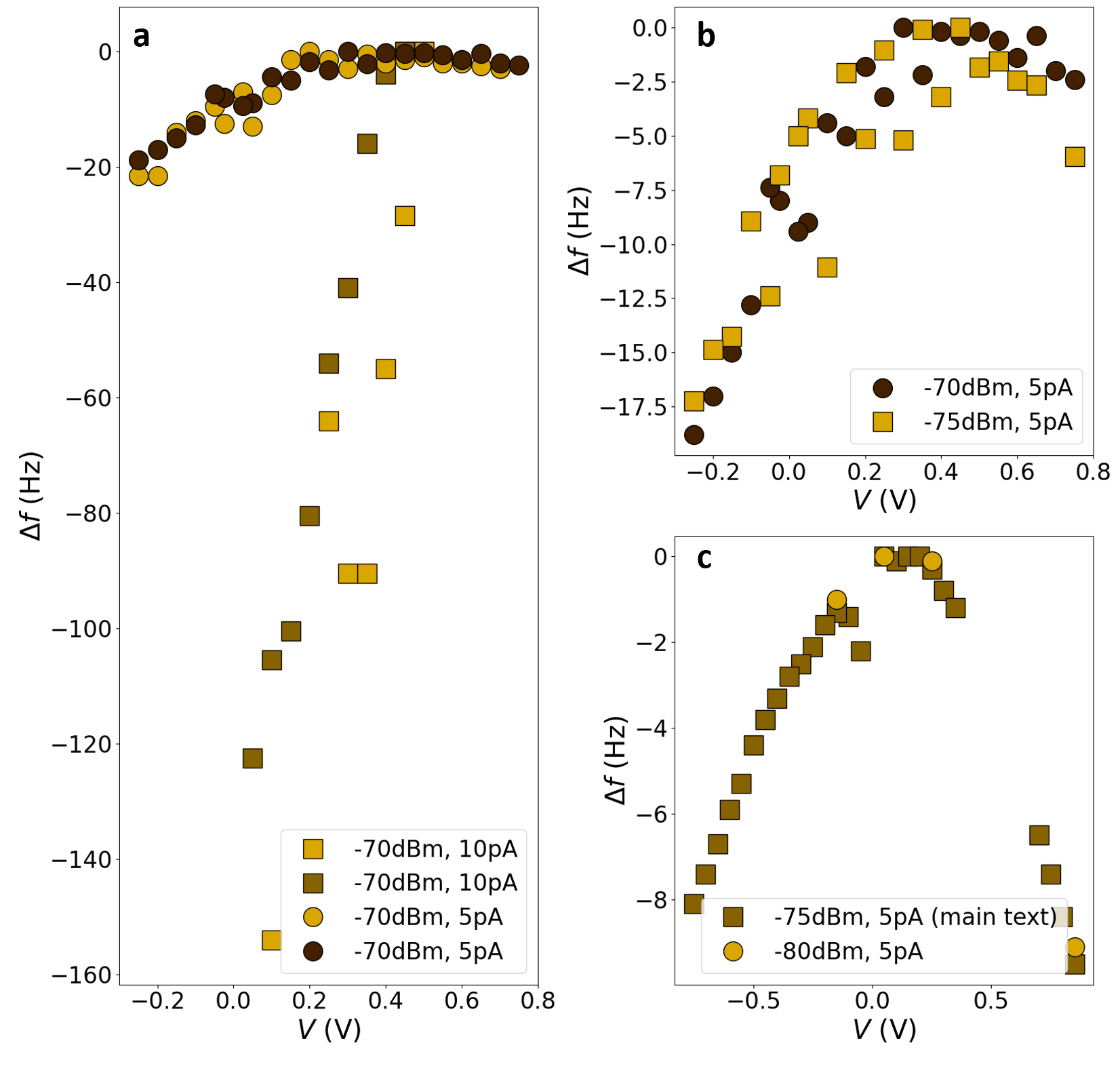}
    \caption{\textbf{LCPD measurements with various different settings} \textbf{(a)} LCPD measurements using Feedback Resonance at two different setpoint currents. Both settings are repeated once. \textbf{(b)} LCPD measurements taken at two different power settings. \textbf{(c)} LCPD measurements also taken at two different power settings, but with a different tip apex than in panel b.}
    \label{fig:lcpd_var}
\end{figure}

The LCPD measurements shown in Figure 3 of the main text were taken with $I = 5$~pA, $P = -75$~dBm, with the current defined at $V_{\rm DC} = 50$~mV. Here we show the effect of changing these values using Feedback Resonance. Figure S\ref{fig:lcpd_var}a shows that with a larger current, and so the tip closer to the membrane, the frequency shifts as a function of $V_{\rm DC}$ are much larger. This is expected, as the electrostatic force induced by the LCPD will be larger for smaller distances. All four data sets shown are taken with the same tip apex, at the same location. Due to the large frequency shifts, limited data was taken, in favour of the smaller frequency shifts at larger tip-membrane distances. Panels b and c show two situations where an LCPD curve is measured with two different driving powers. The dark circles in panel b correspond to the dark circles in panel a. The squares in panel c correspond to the data in the main text. We did not observe any change in the behaviour when varying the driving power, consistent with expectations. Note that the LCPD peak in panel c is closer to $0$~V than the one in panel b. We found the $V_{\rm LCPD}$ to be dependent on the tip apex. Starting with the LCPD curve found in panel b, the tip was purposefully crashed into the surface many times until the curve shown in panel c was found. This curve was preferred as the peak position closer to $0$~V ensured a more stable measurement, see Supplementary Note 7.

\section{Part 9: Approximating $\langle I(t) \rangle$}

Here we will derive an approximation of the current measured during a full membrane oscillation (i.e $1/f_{\rm drive}$, in order to better understand the measured current. We will make the follow assumptions:

\begin{itemize}
    \item \textbf{On-resonance:} we assume that the drive frequency $f_{\rm drive}$ matches exactly the resonance frequency $f_0$ and no off-resonance behavior occurs.
    \item \textbf{Negligible power:} we will assume $V_{\rm DC} \gg V_{\rm RF}$ such that we can assume a constant voltage of $V_{\rm DC}$ for the purposes of integrating $I(t) = G(t) \times V_{\rm DC}$ where $G(t)$ is the tip-membrane conductance. In other words, we disregard the Homodyne component.
    \item \textbf{Equilibrium:} the calculation is based on having driven the membrane long enough that the mechanical amplitude is the same during both the driving (unchopped) and relaxation (chopped). This, together with the previous assumption, allows us to consider just one mechanical oscillation cycle.
    \item \textbf{Exponential conductance dependence:} We will assume, as is typical for STM, the conductance between the tip and membrane has an exponential relation, i.e. $G = G_0 e^{-(d+r)/z_c}$ where $d$ is the distance between the membrane's equilibrium position and the tip, $r$ is the membrane's downwards extension at the tip's position and $d+r$ is the actual distance between the tip and membrane, $z_c$ is some decay length and $G_0$ is the conductance at contact. 
    \item \textbf{Oscillation:} The membrane's extension at the tip's position $r$ follows a sinusoidal function $r = A \sin(2 \pi f_{\rm drive}t)$ where $A$ is the mechanical amplitude at the tip's position. 
\end{itemize}

The average current measured during a lock-in cycle can then be calculated as:

\begin{equation}\label{eq:AvgCur1}
\begin{split}
    \langle I(t) \rangle = f_{\rm drive} \int_0^{\frac{1}{f_{\rm drive}}} G(t)V_{\rm DC}dt \\
    = f_{\rm drive} V_{\rm DC} \int_0^{\frac{1}{f_{\rm drive}}} G_0 e^{-(\frac{d + A\sin(2 \pi f_{\rm drive}t)}{z_c})}  dt \\
    = f_{\rm drive} V_{\rm DC} G_0 e^{-\frac{d}{z_c}}\int_0^{\frac{1}{f_{\rm drive}}} e^{-\frac{A}{z_c}\sin(2 \pi f_{\rm drive}t)} dt \\
    = f_{\rm drive} V_{\rm DC} G_0 e^{-\frac{d}{z_c}}\int_0^{\frac{1}{f_{\rm drive}}} e^{\frac{A}{z_c}\cos(2 \pi f_{\rm drive}t)} dt
\end{split}
\end{equation}

where the last equation holds because the integral is taken over one whole period. We will apply the Jacobi-Anger expansion:

\begin{equation}\label{eq:JacAng}
    e^{iz\cos(\theta)} = \sum_{n=-\infty}^{\infty} i^n J_n(z) e^{in\theta}
\end{equation}

where $J_n(z)$ is the Bessel function of the first kind which accepts complex numbers and $i$ is the imaginary unit $i^2 = -1$. In our case, the complex variable $z = -i\frac{A}{z_c}$, where $A$ and $z_c$ are constants. Plugging this into Eq.~\ref{eq:AvgCur1} results in:

\begin{equation}\label{eq:AvgCur2}
\begin{split}
    f_{\rm drive} V_{\rm DC} G_0 e^{-\frac{d}{z_c}} \int_0^{\frac{1}{f_{\rm drive}}} \sum_{n=-\infty}^{\infty} i^n J_n(-i\frac{A}{z_c}) e^{2\pi i n f_{\rm drive}t} dt \\
    = f_{\rm drive} V_{\rm DC} G_0 e^{-\frac{d}{z_c}} \sum_{n=-\infty}^{\infty} \int_0^{\frac{1}{f_{\rm drive}}} i^n J_n(-i\frac{A}{z_c}) e^{2\pi i n f_{\rm drive}t} dt\\
\end{split}
\end{equation}

Using Euler's formula $e^{ix} = \cos(x) + i \sin(x)$. Any sine or cosine integrated over exactly one period will average out to zero. Thus, for all $n \neq 0$ the integral vanishes, leaving only the following:

\begin{equation}\label{eq:AvgCur3}
\begin{split}
    f_{\rm drive} V_{\rm DC} G_0 e^{-\frac{d}{z_c}} \int_0^{\frac{1}{f_{\rm drive}}} J_0(-i\frac{A}{z_c}) dt \\
    = V_{\rm DC} G_0 e^{-\frac{d}{z_c}} J_0(-i\frac{A}{z_c})
\end{split}
\end{equation}

The Bessel function of the first kind $J_n(x)$ is related to the modified Bessel function of the first kind by $I_n(x) = i^{-n}J_n(ix)$. This means Eq.~\ref{eq:AvgCur3} can be transformed to

\begin{equation}\label{eq:AvgCur4}
\begin{split}
    \langle I(t) \rangle = V_{\rm DC} G_0 e^{-\frac{d}{z_c}} I_0(-\frac{A}{z_c}) \\ 
    = V_{\rm DC} G_0 e^{-\frac{d}{z_c}} I_0(\frac{A}{z_c})
\end{split}
\end{equation}

where the last equation holds due to the symmetry of $I_0$. Thus, once $V_{\rm DC}$, $d$, $z_c$ and $A$ are known, the expected current can be calculated. Alternatively, once the current is measuring, and $V_{\rm DC}$, $d$ and $z_c$ are known, the membrane amplitude $A$ can easily be calculated. \\

To understand which part of the membrane oscillation contributes most strongly to the averaged tunneling current, it is useful to consider the instantaneous current in phase-coordinates. Going back to equation \ref{eq:AvgCur1} and writing $\phi = 2\pi f_{\rm drive} t$, one period corresponds to $\phi\in[0,2\pi]$ and $r(\phi) = A\sin\phi$. The tip--membrane distance is $d+r(\phi)$ and the instantaneous tunneling current is

\begin{equation}
    I(\phi) \propto \exp\!\left[-\frac{d + A\sin\phi}{z_c}\right].
\end{equation}

The average current over one oscillation can then be written as

\begin{equation}
    \langle I\rangle \propto 
    \frac{1}{2\pi}\int_{0}^{2\pi} 
    \exp\!\left[-\frac{d + A\sin\phi}{z_c}\right]\, d\phi .
\end{equation}

The tip--membrane distance is minimal (and the current maximal) when the displacement $r(\phi)$ is most negative, i.e.\ when $\sin\phi$ reaches $-1$.  This occurs at the phase 

\begin{equation}
    \phi_0 = -\frac{\pi}{2} \qquad (\text{equivalently } \phi_0 = 3\pi/2).
\end{equation}

To analyze the contribution of this region, we perform a Taylor expansion around $\phi=\phi_0+\delta$ with $|\delta|\ll 1$.  Using 

\begin{equation}
    \sin(\phi_0+\delta)
    = -\cos\delta
    \approx -\left(1 - \frac{\delta^2}{2}\right)
    = -1 + \frac{\delta^2}{2},
\end{equation}

the exponent becomes

\begin{equation}
\begin{split}
    -\frac{d + A\sin(\phi_0+\delta)}{z_c}
    &\approx -\frac{d + A(-1+\delta^2/2)}{z_c} \\
    &= -\frac{d-A}{z_c} - \frac{A}{2 z_c}\,\delta^2 .
\end{split}
\end{equation}

Thus, near the point of closest approach, the current has the approximate form

\begin{equation}
    I(\phi) \;\approx\;
    I_{\max}
    \exp\!\left[-\frac{A}{2 z_c}\,\delta^2\right],
    \qquad
    I_{\max}\propto \exp\!\left[-\frac{d-A}{z_c}\right],
\end{equation}

i.e.\ it is a narrow Gaussian peak in phase.  
The width of this peak is characterized by

\begin{equation}
    \Delta\phi \;\sim\; 
    \sqrt{\frac{2 z_c}{A}} .
\end{equation}

Because a full oscillation corresponds to $2\pi$ in phase, the fraction of the cycle during which the current is appreciable is therefore

\begin{equation}
    \text{effective duty cycle}
    \;\sim\; 
    \frac{\Delta\phi}{2\pi}
    \sim 
    \frac{1}{2\pi}\sqrt{\frac{2 z_c}{A}}
    \;=\;
    \mathcal{O}\!\left(\sqrt{\frac{z_c}{A}}\right).
\end{equation}

For experimentally relevant amplitudes $A \gg z_c$ this duty cycle is much smaller than unity: the instantaneous current is significant only within a small phase window around $\phi_0$, while during most of the oscillation the membrane is farther from the tip and the current is exponentially suppressed. This demonstrates that the averaged tunneling current arises predominantly from the short portion of the cycle in which the membrane is maximally extended  toward the tip.

\section{Part 10: Further details on extracting $z_{\rm rise}$}
\begin{figure}[ht]
    \centering
    \includegraphics[width=\textwidth]{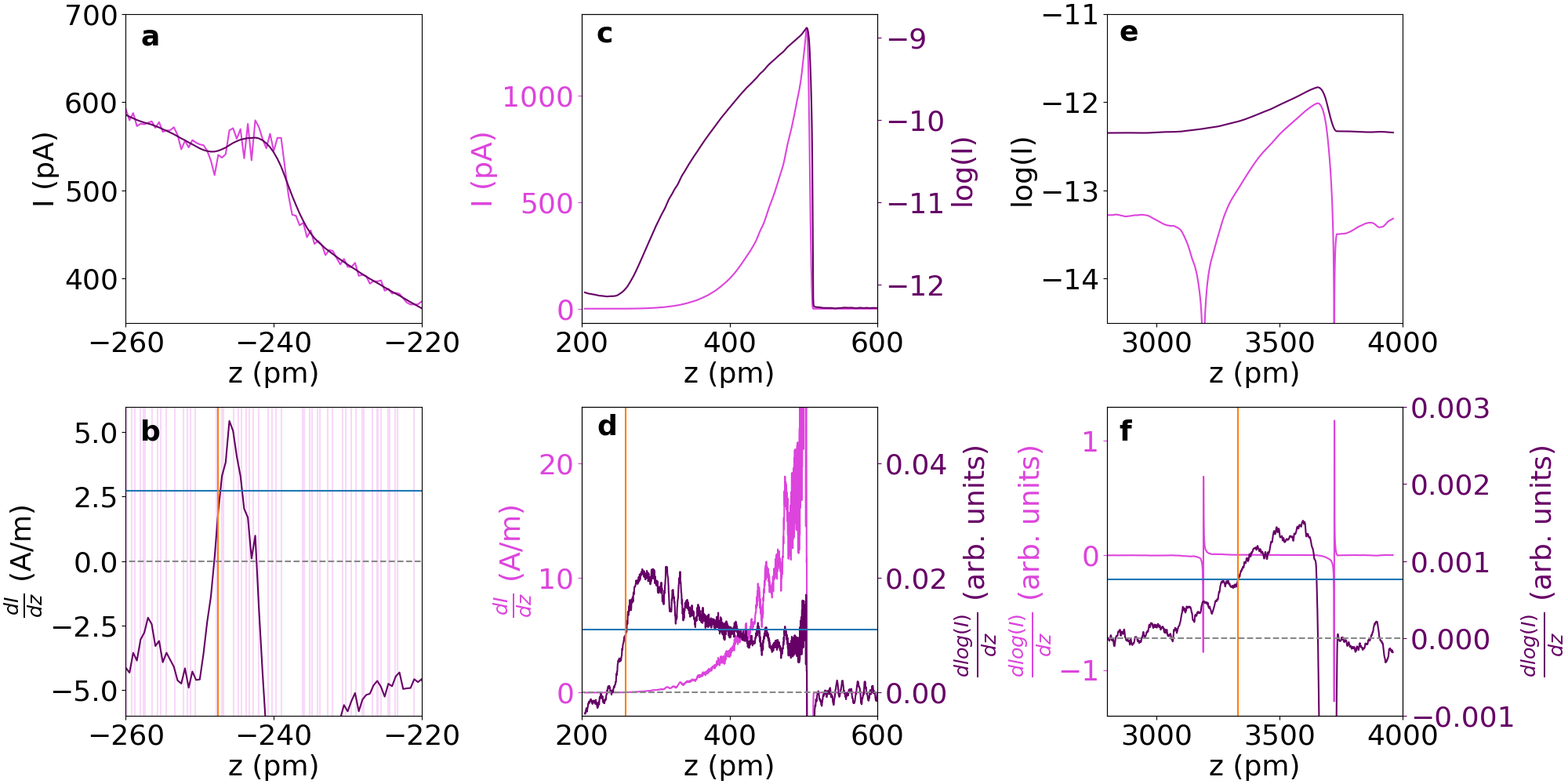}
    \caption{\textbf{Overview of algorithm development.} \textbf{(a)} Raw data (pink) and averaged data with a rolling 15-point triangular weight (purple). \textbf{(b)} Numerical derivative of panel \textbf{a}. Orange and blue lines indicate coordinates where numerical derivative is half its maximum value. \textbf{(c)} Pink and purple data smoothed with rolling 15-point tiraingular weight. Purple data (right axis) shows logarithm of absolute current. \textbf{(d)} Similar to \textbf{b}, but for data shown in panel \textbf{c}. \textbf{(e)} Pink and purple data show smoothed data with the logarithm taken afterwards. Prior to smoothing, purple data is offset by +50 fA. \textbf{(f)} similar to \textbf{b}, but for data shown in panel \textbf{e}. }
    \label{fig:algorithm}
\end{figure}

We did not use the exact same algorithm for determining $z_{\rm rise}$ throughout the entire paper. This is due to different noise levels, different power levels and different behavior depending on whether $z_{\rm rise}$ occurs at the slope of decreasing $I$ or in the noise when $z$ is large enough and we rely on $\tilde{z}_{\rm rise}$ instead. Overall, a combination of three processes are used: smoothing the data, taking the logarithm of the data and adding some offset to the data. After any or all of these processes, the derivative of the current with respect to the tip-sample distance is taken, and the point where the derivative reaches half the maximum value is taken as $z_{\rm rise}$ (or $\tilde{z}_{\rm rise}$). Figure \ref{fig:algorithm} highlights these processes in detail.

Panels a, c and e show the current of three different measurements whereas panels b, d and f show the derivative of the current, as well as the numerically extracted value of $z_{\rm rise}$ (orange line) where the derivative reaches half the maximum value of the curve (blue line). \\

In panel a, the pink curve shows the raw current, whereas in purple the data is averaged, using a triangular window of width 15 points centered on the point in question. Panel b shows the derivative of both curves. Only the smoothed data provides a stable derivative for the algorithm to work with.

In panel c, both curves are already smoothed. As the pink curve shows, current has a continuously increasing slope. Then, taking half of the maximum derivative would result in a $z_{\rm rise}$ close to $z_{\rm drop}$. Instead, we first take the logarithm of the current, as shown with the purple curve. As panel d reveals, this puts the maximum derivative of the (log of) the current much closer towards the actual increase in current. Now taking half of the maximum derivative yields a fairly accurate value of $z_{\rm rise}$.

Finally, in panel~e the pink curve shows the current after smoothing and after taking the logarithm of its absolute value. As there was a small negative current offset during the acquisition of this data, taking the logarithm results in singularities whenever the raw data crosses zero. These singularities persist in the derivative, as evidenced in panel~f. To accommodate this, the purple curve in panel~e has an offset of $500$~fA added to the raw data prior to taking the logarithm. This removes the singularities and yields a derivative with a well-defined maximum from which its half can be determined. For numerical stability, offsets of order $10^2$~fA—comparable to the measurement noise floor—were applied where necessary, although the exact offset varies between datasets.

The data presented in Fig.~5b of the main text instead uses $\tilde{z}_{\rm rise}$, defined by the current exceeding a fixed threshold rather than by the derivative reaching half its maximum. In datasets where the physical onset of the current increase occurs well below the noise floor, neither definition yields the true onset distance. In this regime, $\tilde{z}_{\rm rise}$ serves as a detection-limited proxy. For the purpose of producing the histograms in Fig.~5c in a visually intuitive and robust manner, $\tilde{z}_{\rm rise}$ was therefore defined using a threshold of $60$~fA, corresponding to the approximate current noise floor. Operationally, this was implemented by adding $500$~fA to the data, smoothing, taking the logarithm, and applying a fixed cutoff at $-12.25$ in log space.

For more details, please see the Open Data Folder (upload pending),

\section{Part 11: Further details on Figures 4a and 5b}
\begin{figure}[ht]
    \centering
    \includegraphics[width=\textwidth]{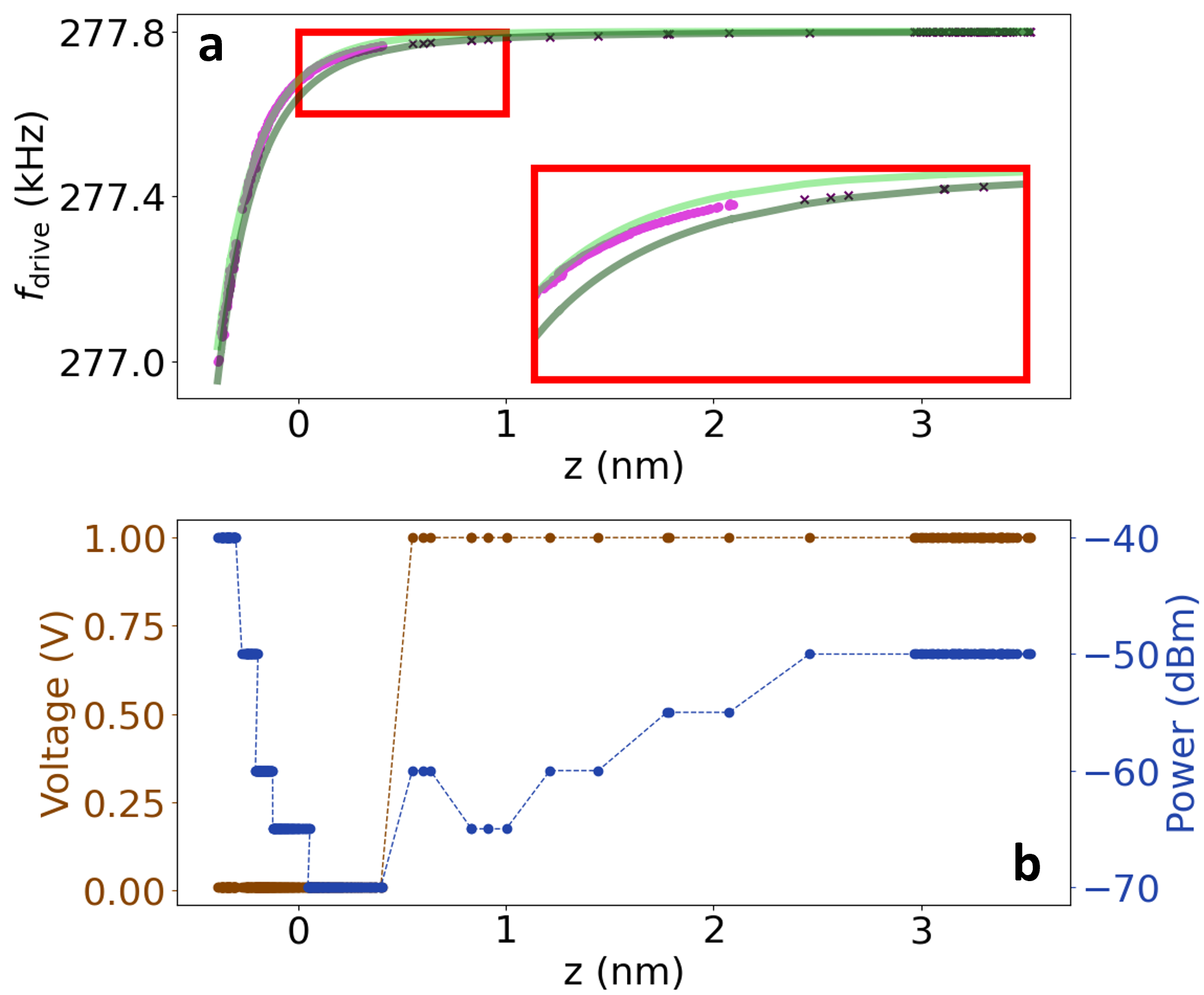}
    \caption{\textbf{Further Details on Figure 5a.} \textbf{(a)} Same data as shown in Figure 5a, but now with a zoom-in on the transition from $V_{\rm DC} = 100~mV$ to 1~V. \textbf{(b)} Applied voltages $V_{\rm DC}$ and applied power $P$ at every tip distance position $z$.}
    \label{fig:LJ_extra}
\end{figure}

The settings used to acquire the data in Figure 4a were not consistent throughout the entire data range. In particular, as descrobed in the main text, two different values for bias voltage $V_{\rm DC}$ were used. Additionally, various values for driving power $P$, as shown in Fig.~\ref{fig:LJ_extra}b. To account for the different voltages, one simply fills in different values for $V_{\rm DC}$ in Eq.~3 of the main text. Meanwhile, the different powers ought not to change the measured $z_{\rm rise}$ much, as evidenced by Figure 2 of the main text. Nonetheless, there is still a finite power dependence where $z_{\rm rise}$ increases (towards the actual $z_0$) for decreasing powers. As a result, for data acquired at lower power (i.e. $z \approx 300$~pm), the values of $z_{\rm rise}$ will be larger than the fit (i.e. data points are to the right of the fitted curve), as shown in the inset of Fig.~\ref{fig:LJ_extra}.\\

Additionally, due to the very high powers in Eq. 1 of the main text, coefficients of significantly different magnitudes and the fact that data spans several orders of magnitude (in terms of detuning), the fitting procedure suffered from numerical accuracy. \\

Both above-mentioned issues put limits on the accuracy of the fit to the model, which limits the accuracy of the force resolution claim. Furthermore, the tip apex slightly changed between taking the data of Fig.~5a and Fig.~5b. We did not conduct a thorough study on the effects of this, but note that $V_{\rm LCPD}$ only changed when the tip apex was altered significantly, see Supplementary Note 8. With this in mind, we expect the actual forces imparted on the membrane by the tip, as measured in Fig.~5b, to be similar to but not exactly equal to what was presented there. To account for this, Table \ref{tab:freqs_changes} considers several different coefficients for for Eq.~2 and shows the calculated force resolution from these settings. We did not consider a change in $f_0$ as this is inherent to the membrane, rather than dependent on the tip. We find that the order of magnitude does not really change.

\begin{table}[]
\begin{tabular}{|l|l|l|}
\hline
\textbf{Parameter}       & \textbf{10\% bigger}          & \textbf{10\% smaller}         \\ \hline
$\epsilon_{\rm LJ}$ & $6.1 \cdot 10^{-12}$ & $6.1 \cdot 10^{-12}$ \\ \hline
$\sigma$        & $6.2 \cdot 10^{-12}$ & $6.1 \cdot 10^{-12}$ \\ \hline
$C_1$           & $6.7 \cdot 10^{-12}$ & $5.5 \cdot 10^{-12}$ \\ \hline
$d_0$           & $5.5 \cdot 10^{-12}$ & $6.8 \cdot 10^{-12}$ \\ \hline
\end{tabular}
\caption{Calculated force resolutions if the fit values were slightly different.}\label{tab:freqs_changes}
\end{table}

\end{document}